\documentclass[journal]{IEEEtran}
\IEEEoverridecommandlockouts

\ifCLASSINFOpdf
\else
   \usepackage{graphicx}
   \graphicspath{{../eps/}}
   \DeclareGraphicsExtensions{.eps}
\fi

\usepackage{amsmath}
\usepackage{amsfonts}
\usepackage{amssymb}
\usepackage{wrapfig}
\usepackage{psfrag}
\usepackage{epstopdf}
\usepackage{cite}
\usepackage{graphicx}
\usepackage{subfigure}
\usepackage{threeparttable}
\usepackage{cases}
\usepackage{subeqnarray}
\usepackage{color}
\usepackage{underscore}
\usepackage[algo2e,ruled,vlined]{algorithm2e}
\usepackage{algorithmic}

\usepackage{textcomp}
\usepackage{xcolor}
\def\BibTeX{{\rm B\kern-.05em{\sc i\kern-.025em b}\kern-.08em
    T\kern-.1667em\lower.7ex\hbox{E}\kern-.125emX}}

\begin{document}
\title{MIMO Symbiotic Radio with Massive\\Passive Devices: Asymptotic Analysis\\and Precoding Optimization}

\author{Jingran~Xu,~Zhuoyin~Dai,~and~Yong~Zeng,~\IEEEmembership{Member,~IEEE}
        % <-this % stops a space
\thanks{This work was supported by the National Key R\&D Program of China with
grant number 2019YFB1803400, by the Natural Science Foundation of China under Grant 62071114, and also by the Fundamental Research Funds for the Central Universities 2242022k30005. Part of this work has been presented at the
2021 IEEE GLOBECOM, Madrid, Spain, 7-11 Dec. 2021 \cite{conference}.}% <-this % stops a space
\thanks{The authors are with the National Mobile Communications Research
Laboratory and Frontiers Science Center for Mobile Information Communication and Security, Southeast University, Nanjing 210096, China. Y. Zeng is also
with the Purple Mountain Laboratories, Nanjing 211111, China (e-mail:
{jingran_xu,~zhuoyin_dai,~yong_zeng}@seu.edu.cn). (\emph{Corresponding author: Yong Zeng.})}}

\maketitle

% \vspace{-0.5cm}
\begin{abstract}
Symbiotic radio has emerged as a promising technology for spectrum- and energy-efficient wireless communications, where the passive secondary backscatter devices (BDs) reuse not only the spectrum but also the power of the active primary users to transmit their own information. In return, the primary communication links can be enhanced by the additional multipaths created by the BDs. This is known as the \emph{mutualism} relationship of symbiotic radio.
However, due to the severe double-fading attenuation of the passive
backscattering links,
%as the passive backscattering links are typically much weaker than the direct link due to double attenuations,
the enhancement of the primary link provided by one single BD is extremely limited. To address this issue and enable full mutualism of symbiotic radio, in this paper, we study multiple-input multiple-output (MIMO) symbiotic radio communication systems with massive BDs. We first derive the achievable rates of the primary active communication and secondary passive communication, and then consider the asymptotic regime as the number of BDs goes large, for which closed-form expressions are derived to reveal the relationship between the primary and secondary communication rates. Furthermore, the precoding optimization problem is studied to
maximize the primary communication rate while guaranteeing
that the secondary communication rate is no smaller than
a certain threshold. Simulation results are provided to validate our theoretical studies.
\end{abstract}


\IEEEpeerreviewmaketitle
% \vspace{-0.3cm}
\section{Introduction}
%As one of the promising technologies of the next generation Internet of things(IoT), symbiotic radio can save spectrum resources and reduce energy consumption effectively~\cite{IEEEhowto:1}, in which backscatter devices(BD) transmit their information over the incident primary signal via backscatter modulation instead of using active radio-frequency (RF) components~\cite{IEEEhowto:2},~\cite{IEEEhowto:3}.\\
Symbiotic radio has emerged as a new paradigm to achieve both spectrum-efficient and energy-efficient wireless communications, for which the secondary user modulates its information over the radio frequency (RF) signals received from primary transmitter (PT) \cite{applications,IEEEhowto:2,IEEEhowto:3}.
%to achieve both spectrum- and energy-efficient wireless communications \cite{applications,IEEEhowto:2,IEEEhowto:3,IEEEhowto:6G}.
%promising candidate for sixth-generation (6G)
%For typical symbiotic radio systems, the secondary user modulates information symbols over ambient radio frequency (RF) signals received from a primary transmitter (PT).
As such, the secondary backscatter device (BD) in symbiotic radio systems leverages not only the spectrum as in the extensively studied cognitive radio systems \cite{IEEEhowto:4,IEEEhowto:5,IEEEhowto:6}, but also the energy of the primary signals via passive backscattering technology for its own information transmission.
%Different from the extensively studied cognitive radio systems \cite{IEEEhowto:4,IEEEhowto:5,IEEEhowto:6}, the secondary backscatter device (BD) in symbiotic radio systems leverages not only the spectrum but also the power of the primary signals for its own information transmission.
To overcome the shortcoming of poor reliability suffered from the conventional passive ambient
backscatter communication (AmBC) receiver \cite{AmBC1,AmBC2,AmBC3}, symbiotic radio introduces cooperation between the backscatter transmission and the primary transmission by using a joint receiver.
Depending on the relations between the symbol durations of the primary and secondary signals, symbiotic radio can be classified into \emph{parasite symbiotic radio} (PSR) and \emph{commensal symbiotic radio} (CSR) \cite{R.Long}. In PSR, the secondary and primary signals have equal symbol durations, so that the information transmission of BD interferes with the primary transmission. By contrast, for CSR, the symbol duration of BD signals is much longer than that of the primary signals, so that the backscattering signal of the BD may create additional multipaths to enhance the primary communication links. This is known as the \emph{mutualism} relationship of symbiotic radio \cite{R.Long}. The mutualism spectrum sharing and low power consumption nature of symbiotic radio render it an attractive massive access technology for the sixth-generation (6G) mobile communication networks \cite{massiveaccess1,massiveaccess2,massiveaccess3}, to realize spectrum- and energy-efficient Internet of Things (IoT).
%The mutualism spectrum sharing and low power consumption nature of symbiotic radio make it suitable as a massive access technology for 6G and beyond \cite{massiveaccess1,massiveaccess2,massiveaccess3} to achieve spectrum- and energy-efficient Internet of Things (IoT) networks.
By riding over different types of primary networks, there are many emerging applications for symbiotic radio, such as e-health, smart home, and environmental monitoring \cite{applications}.

Significant research efforts have been recently devoted to the theoretical analysis and performance optimization of symbiotic radio systems. For example, the performance analysis in terms of achievable rate \cite{achievablerate1,achievablerate2}, channel capacity \cite{capacity} and outage probability \cite{NOMA1} are given in different setups.
%The practical receiver design of symbiotic radio systems is considered in \cite{IEEEhowto:11,IEEEhowto:12,IEEEhowto:13}.
By exploiting multi-antenna techniques for performance enhancement, multiple-input-single-output (MISO)
and multiple-input-multiple-output (MIMO) symbiotic radio systems are investigated in \cite{achievablerate1,IEEEhowto:14,IEEEhowto:15}.
Furthermore, beamforming optimization problems have been studied to maximize various performance metrics, such as energy efficiency \cite{achievablerate1}, sum capacity of the primary and secondary communications \cite{IEEEhowto:14}, and fairness of the secondary users \cite{IEEEhowto:15}.
Besides, energy efficiency (EE) studies for symbiotic radio systems have also received growing attentions recently \cite{EE1,EE2,EE3}. For example, in
\cite{EE1} and \cite{EE2}, the EE of symbiotic radio is defined as the ratio of the total throughput of all links to the total energy consumption of the network, while \cite{EE3} characterizes the EE region of symbiotic radio systems, which is defined as all the achievable EE pairs by the active PT and passive BD.
In addition, the combination of symbiotic radio with other technologies has also been investigated. For example, systems integrating non-orthogonal-multiple-access
(NOMA) into symbiotic radio have been studied in \cite{NOMA1,NOMA2,NOMA3}. In \cite{fullduplex1,fullduplex2}, full-duplex technique is introduced into a symbiotic radio system, which enables a BD to transmit and receive information simultaneously. In \cite{cellfree}, cell-free networking architecture is integrated with symbiotic radio transmission technology to realize passive communication with high macro-diversity.

%Despite of the potential benefits, symbiotic radio also faces new critical challenges.
Note that due to the severe double-fading attenuation, the strength of the backscattering link
is usually much weaker than that of the direct link, which
results in low data rates for the secondary communication
and only marginal enhancement to the primary transmission by the backscattering link in the CSR setup.
%In particular, as the backscattered signal suffers from double power attenuation, the strength of the backscattering link is usually much weaker than that of the direct link between the primary transmitter (PT) and primary receiver.
%This not only limits the communication performance of the secondary system itself, but also compromises its promised performance enhancement to the primary communication links.
To address such issues, various techniques have been proposed, such as the active-load assisted \cite{activeload1,activeload2,activeload3} symbiotic radio by using negative resistances,  or reconfigurable intelligent surface (RIS) aided \cite{RIS1,RIS2,RIS3} symbiotic radio by configuring large scale passive reflecting elements.
However, the use of active loads requires additional
power supply to generate the negative resistances, and a large scale deployment in RIS may not be suitable for many IoT devices that have small form factors. Thus, such existing techniques for enhancing the secondary communication links may increase the cost and complexity of the passive BD, which undermines the initial motivation of symbiotic radio.
%However, such techniques increase the complexity of the structure of traditional BD and may drastically increase the power, cost and size of it, which may undermine the initial motivation of symbiotic radio.

In this paper, we propose an alternative method to significantly enhance the secondary backscattering links and enable the full mutualism of symbiotic radio systems, by exploiting the potential gain brought by multiple BDs, which provides abundant multi-user diversities for secondary communications and multipath gains for primary transmission. This is motivated by the 6G visions to support ultra-massive connectivity, say 10 million devices per square kilometer \cite{IEEEhowto:20}, most of which are expected to be IoT devices.
%This thus motivates our current work to study symbiotic radio with massive BDs, which, to the best of
%our knowledge, have not been studied in the existing literature.
%As such, both the secondary passive and the primary active communication rates can be enhanced by accessing massive BDs due to multi-user diversity gains and multipath gains.
Our main contributions are summarized as follows:
\begin{itemize}
	\item First, we present the mathematical model of MIMO symbiotic
radio communication systems with multiple BDs in the CSR setup, where the mutualism relationship of symbiotic
radio can be fully exploited.
We derive the achievable rate expression of the primary communication, as well as the sum rate expression of the passive BDs, by noting that it corresponds to a multiple access channel (MAC) where minimum mean square error estimation (MMSE) with successive interference cancellation (SIC) is optimal \cite{IEEEhowto:21}.	

\item Next, to show how the achievable communication rates of symbiotic radio are affected by the number of BDs, we consider the asymptotic analysis as the number of BDs goes large. Closed-form expressions are derived for the primary active communication rate and secondary passive communication rate, both of which are shown to be increasing functions of the number of BDs.
For the special case of single-input-multiple-output (SIMO) symbiotic radio setup, we derive the closed-form expression of the asymptotic primary communication rate as a function
of the asymptotic secondary sum-rate, which is revealed to be an increasing function.
This thus demonstrates that the mutualism relationship of symbiotic radio can be fully enabled with massive BD access.

\item  Furthermore,
we formulate a precoding optimization problem to maximize the primary communication rate by taking into account the additional multipaths created by the BDs, while guaranteeing that the secondary communication rate is
no smaller than a certain threshold.
The problem is non-convex in its original form, and we propose an effective technique to transform it to an equivalent convex problem of optimizing the transmit covariance matrix. Two solutions with different trade-offs between complexity and performance are proposed, namely \emph{sample-average based solution} and \emph{upper bound based solution}.
Numerical results are provided to demonstrate that
the proposed optimization approaches are effective
in MIMO symbiotic radio communication systems.

\end{itemize}

The rest of this paper is organized as follows. Section II
presents the system model of MIMO symbiotic
radio communication with multiple BDs.
Section III derives the achievable rate of the primary communication and the sum rate of the secondary BDs.
In Section IV, by assuming that the number of BDs goes large, asymptotic performance analysis is provided to reveal the relationship between primary active communication and secondary passive communication.
In Section V, the precoding optimization problem is studied to maximize the primary communication rate while guaranteeing that the secondary communication rate is no smaller than a certain threshold.
Section VI presents numerical results to validate our theoretical studies.
Finally, we conclude the paper in Section VII.

\emph{Notations:} In this paper, scalars are denoted by italic letters.
Vectors and matrices are denoted by boldface lower-and upper-case letters,
respectively. For a vector $\mathbf{a}$, its
transpose, Hermitian transpose, and Euclidean norm are respectively
denoted as $\mathbf{a}^{\text T}$, $\mathbf{a}^{\text H}$ and $\| \mathbf{a}\|$.
The conjugate, transpose, Hermitian transpose and rank of a matrix $\mathbf{A}$ are denoted as
${{\mathbf{A}}^{*}}$, ${{\mathbf{A}}^{\text{T}}}$, ${{\mathbf{A}}^{\text{H}}}$ and $\text{rank}(\mathbf{A})$, respectively.
$\text{vec}(\mathbf{A})$ represents stacking the columns of matrix $\mathbf{A}$ into a column vector.
For a square matrix $\mathbf{B}$, ${\rm tr}(\mathbf{B})$, $\left| \mathbf{B} \right|$, ${{\mathbf{B}}^{-1}}$, and ${{\mathbf{B}}^{\frac{1}{2}}}$
denote its trace, determinant, inverse, and matrix square-root, respectively, while $\mathbf{B}\succeq0$ represents that the matrix $\mathbf{B}$ is positive semidefinite.
${{\lambda }_{1}(\mathbf{B})},...,{{\lambda }_{M}(\mathbf{B})}$ denote the eigenvalues of $\mathbf{B}$.
$\otimes$ refers to the Kronecker product.
diag$(x_1,...,x_M )$ denotes an
$M\times M$ diagonal matrix with $x_1,...,x_M $ being the diagonal elements.
$\mathbf I_M$ denotes an $M \times M$ identity matrix.
$\mathbb{C}^{M\times N}$ denotes the space of ${M\times N}$ matrices with complex entries.
$\mathbb{E}_{X}[\cdot]$ denotes the expectation with respect to the
random variable $X$.
$\log_{2}(\cdot) $ denotes the logarithm with base 2. Furthermore,
$\mathcal{CN}(\mathbf{x},\mathbf{\Sigma})$ denotes the
distribution of a circularly symmetric complex Gaussian (CSCG) random vector with mean $\mathbf{x}$ and covariance matrix $\mathbf{\Sigma}$.

\section{System Model}
\begin{figure}[!t]
  \centering
  \centerline{\includegraphics[width=3.5in,height=2.0in]{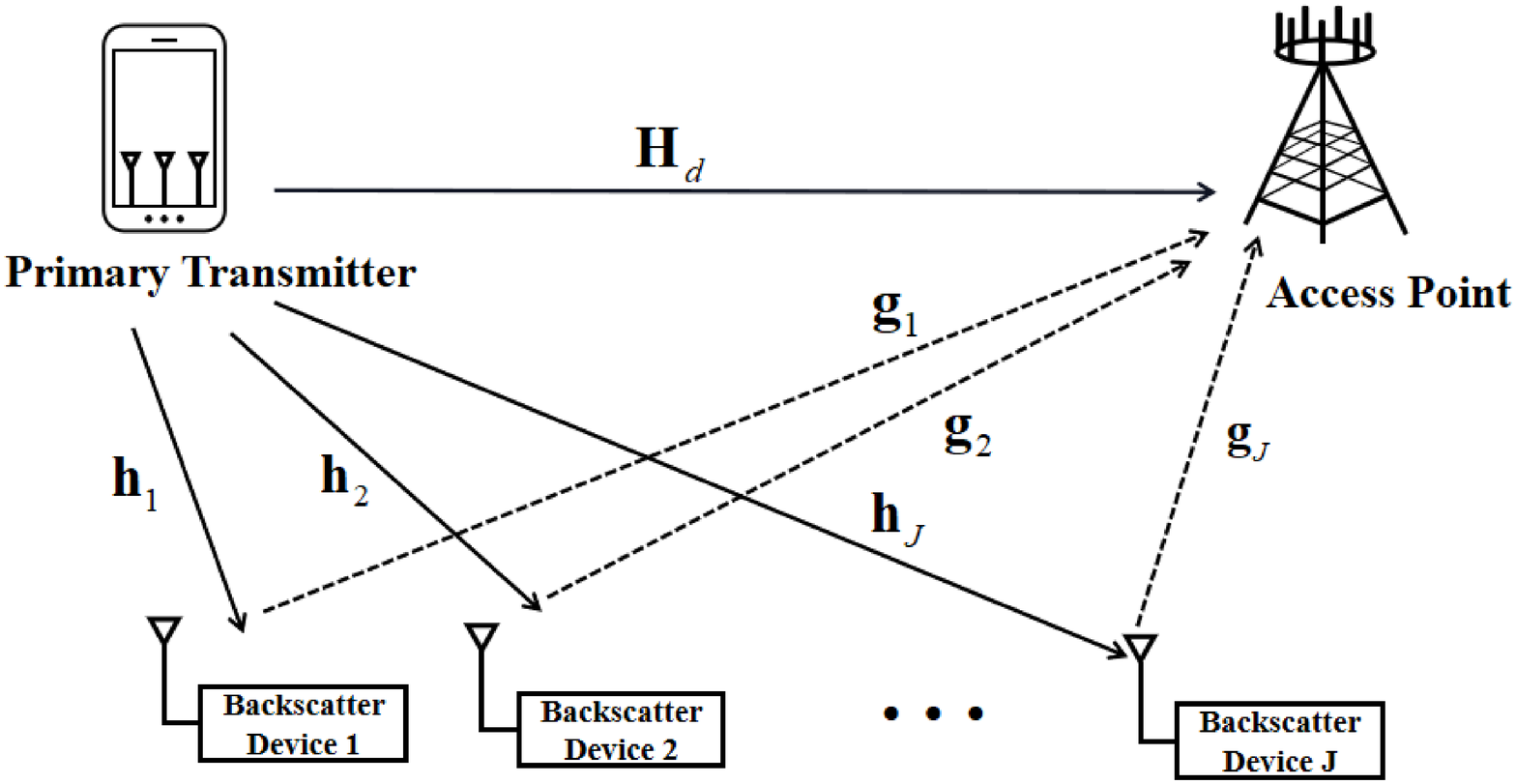}}
  \caption{MIMO symbiotic radio with massive backscatter devices.}
  \label{system model}
   \vspace{-0.5cm}
  \end{figure}
As shown in Fig. 1, we consider a MIMO symbiotic radio communication system, which consists of one PT, one access point (AP) and $J$ BDs. We assume that the PT and AP have $M_t$ and $M_r$ antennas, respectively,  and each BD is equipped with one antenna. The AP wishes to decode not only the primary information from the PT, but also the secondary information from the $J$ BDs, which modulate their information via passively backscattering the incident signal from the PT. As such, the BDs reuse not only the spectrum but also the power of the PT. In return, their scattered signals may create additional channel paths to enhance the primary communication link, as long as their symbol rate is much lower than that of the primary signal. This is known as \emph{mutualism} relationship of symbiotic radio\cite{R.Long}.
Denote the MIMO direct-link channel from the PT to the AP as ${{\mathbf{H}}_{d}}\in {{\mathbb{C}}^{M_r\times M_t}}$. Further denote by  ${\mathbf{h}_{j}}\in \mathbb{C}^{M_t\times 1}$ the MISO channel from the PT to BD $j$, and $\mathbf g_j\in \mathbb{C}^{M_r\times 1}$ the SIMO channel from BD $j$ to the AP, where $j=1,...,J$. Then the channel matrix for the cascaded backscattering link from the PT to AP via BD $j$ is $\mathbf g_j\mathbf{h}_{j}^{\text{H}}$.

We focus on the CSR setup \cite{R.Long}, where the symbol duration of the BDs is $K\gg 1$ times of that of the PT. Let $c_j(n)$ denote the independent and identically distributed (i.i.d.) information-bearing symbols of BD $j$, and $\mathbf{s}(k,n)\in \mathbb{C}^{M_s \times 1}$ denote the i.i.d. information-bearing symbols of the PT, which follows CSCG distribution with normalized power, i.e., $\mathbf s(k,n)\sim\mathcal{C}\mathcal{N}(\boldsymbol 0, \mathbf I_{M_s})$, $k=1,...,K$. Note that $M_s\leq M_t$ denotes the number of data streams of the PT signal, which is one of the optimization variables to be determined later.
%We assume that $\mathbf{s}(k,n)$ follows the independent and identically distributed (i.i.d.) circularly symmetric complex gaussian (CSCG) distribution, i.e., $s(k,n)\sim \mathcal{C}\mathcal{N}(0,1)$.
Furthermore, let $P$ denote the transmit power by the PT, $\alpha \in \left[ 0,1 \right]$ be the fraction of the power backscattered by each BD, and $\mathbf{F}\in {{\mathbb{C}}^{M_t\times M_s}}$ denote the transmit precoding matrix of the PT, with $\text{tr}(\mathbf{F}{{\mathbf{F}}^{\text{H}}})=1$.
Then the signal received by the AP during the $n$th BD symbol duration is
\begin{equation}
\setlength\abovedisplayskip{1pt}
\setlength\belowdisplayskip{1pt}
 \begin{aligned}
\mathbf{y}(k,n)=\sqrt{P}{{\mathbf{H}}_{d}}\mathbf{Fs}(k,n)+\sum\limits_{j=1}^{J}{\sqrt{P}}\sqrt{\alpha }{{\mathbf{g}}_{j}}{{\mathbf{h}}_{j}^\text{H}}\mathbf{Fs}(k,n){{c}_{j}}(n)\\
+\mathbf{z}(k,n),k=1,...,K,\label{eq:origin}
 \end{aligned}
\end{equation}
where $\mathbf{z}(k,n)\in \mathbb{C}^{M_r\times 1}$ is the i.i.d. CSCG noise with zero mean and power ${{\sigma }^{2}}$, i.e., $\mathbf{z}(k,n)\sim\mathcal{C}\mathcal{N}\left( \mathbf 0, \sigma^2{\mathbf{I}}_{M_r}\right)$.

% >>>>>>>>>>>>>SECTIONS II -  here >>>>>>>>>>>>
%\vspace{-0.1cm}
\section{Achievable Rate Analysis}
Since the BD symbols $c_j(n)$ remain unchanged for each block of $K$ PT symbols, the second term in \eqref{eq:origin} constitutes additional multi-path channel components for the primary signal. As a result, the equivalent MIMO channel for decoding $\mathbf{s}(k,n)\in \mathbb{C}^{M_s\times 1}$ is dependent on the BD symbols $\mathbf c(n)=[c_1(n),c_2(n),...,c_J(n)]^{\text{T}}$, which is denoted as ${{\mathbf{H}}_{eq}}(\mathbf{c}(n))={{\mathbf{H}}_{d}}+\sum\limits_{j=1}^{J}{\sqrt{\alpha }}{{\mathbf{g}}_{j}}\mathbf{h}_{j}^{\text{H}}{{c}_{j}}(n)$. Therefore, \eqref{eq:origin} can also be expressed as
\begin{equation}
{\mathbf{y}(k,n)=\sqrt{P}{{\mathbf{H}}_{eq}}(\mathbf{c}(n))\mathbf{Fs}(k,n)+\mathbf{z}(k,n)},k=1,...,K.\label{eq:equivalent}
\end{equation}
Note that \eqref{eq:equivalent} is essentially a block fading channel where the MIMO channel matrix ${\mathbf{H}_{eq}(\mathbf{c}(n))}$ remains unchanged in each block $n$ of $K$ symbol durations, while varies across different blocks according to the BD symbols $\mathbf{c}(n)$.
We assume that the AP has the knowledge of the equivalent MIMO channel ${\mathbf{H}}_{eq}(\mathbf{c}(n))$, e.g., via standard training-based MIMO channel estimation \cite{howmuchtraining}, while the PT only knows the information of ${\mathbf{H}}_{d}$, $\mathbf{h}_{j}$ and $\mathbf g_j$, but not ${\mathbf{H}}_{eq}(\mathbf{c}(n))$, since the BD symbols $\mathbf{c}(n)$ are unknown at the PT.
In this case, with a sufficiently large $K$, the average primary communication rate for the input-output relationship \eqref{eq:equivalent} is
\begin{equation}
\begin{split}
{{R}_{s}}={{\mathbb{E}}_{\mathbf{c}(n)}}\left[ {{\log }_{2}}\left| {{\mathbf{I}}_{M_r}}+{\bar P}{{\mathbf{H}}_{eq}}\left( \mathbf{c}\left( n \right) \right)\mathbf{F}{{\mathbf{F}}^{\text{H}}}\mathbf{H}_{eq}^{\text{H}}\left( \mathbf{c}\left( n \right) \right) \right| \right],\label{eq:primaryrate}
\end{split}
\end{equation}
where ${\bar P}\triangleq\frac{P}{{{\sigma }^{2}}}$ is defined as the transmit signal-to-noise ratio (SNR).

On the other hand, to decode the symbols $c_j(n)$ for each BD $j$, by concatenating $\mathbf{y}(k,n)$ in \eqref{eq:origin} for all $k=1,2,\cdots ,K$, we have
$\mathbf{Y}(n)=\left[ \mathbf{y}(1,n),\mathbf{y}(2,n),\cdots ,\mathbf{y}(K,n) \right]\in \mathbb{C}^{M_r\times K}$.
Similarly, let $\mathbf{S}(n)={\left[ {\mathbf{s}}{(1,n)},{\mathbf{s}}{(2,n)},\cdots ,{\mathbf{s}}{(K,n)}\right]\in  \mathbb{C}^{M_s\times K} }$ and $\mathbf{Z}(n)=\left[ \mathbf{z}(1,n), \mathbf{z}(2,n),\cdots ,\mathbf{z}(K,n) \right]\in  \mathbb{C}^{M_r\times K}$.
Then \eqref{eq:origin} can be compactly written as
\begin{equation}
{\mathbf{Y}(n)\!=\!\sqrt{P}{{\mathbf{H}}_{d}}\mathbf{F}\mathbf{S}(n)\!+\!\sum\limits_{j=1}^{J}{\sqrt{P}}\sqrt{\alpha }{{\mathbf{g}}_{j}}\mathbf{h}_{j}^{\text{H}}\mathbf{F}\mathbf{S}(n){{c}_{j}}(n)\!+\!\mathbf{Z}(n)}.\label{eq:compacteq}
\end{equation}
After decoding $\mathbf{s}(k,n)$ at the AP, $k=1,...,K$, the primary signal component $\mathbf{S}(n)$ can be subtracted from \eqref{eq:compacteq} before decoding the BD signals, which yields
\begin{equation}
{\mathbf{\hat{Y}}(n)=\sum\limits_{j=1}^{J}{\sqrt{P}}\sqrt{\alpha }{{\mathbf{g}}_{j}}\mathbf{h}_{j}^{\text{H}}\mathbf{F}\mathbf{S}(n){{c}_{j}}(n)+\mathbf{Z}(n)}.\label{eq:subtractps}
 \end{equation}
Furthermore, with $\mathbf{S}(n)$ decoded at the receiver, the temporal-domain matched filtering can be applied, by right multiplying $\mathbf{\hat{Y}}(n)$ in \eqref{eq:subtractps} with $(1/\sqrt{K})\mathbf{S}^{\text{H}}{(n)}$.
For sufficiently large $K$, due to the law of large numbers and the fact that the information-bearing symbols $\mathbf s(k,n)$ are i.i.d. with distribution $\mathbf s(k,n)\sim\mathcal{C}\mathcal{N}(\boldsymbol 0, \mathbf I_{M_s})$, we have $(1/K)\mathbf{S}(n)\mathbf{S}^{\text{H}}{{(n)}}\to {{\mathbf{I}}_{M_s}}$.
Therefore, the resulting signal is
\begin{equation}
\begin{split}
\mathbf{\tilde{Y}}(n)& =\frac{1}{\sqrt{K}}\mathbf{\hat{Y}}(n)\mathbf{S}^{\text{H}}{{(n)}}\\
& =\sum\limits_{j=1}^{J}{\sqrt{KP\alpha }}{{\mathbf{g}}_{j}}\mathbf{h}_{j}^{\text{H}}\mathbf{F}{{c}_{j}}(n)+\mathbf{\tilde{Z}}(n),\label{eq:tbbeam}
\end{split}
\end{equation}
where $\mathbf{\tilde{Z}}(n)\!=\!\frac{1}{\sqrt{K}}\mathbf{Z}(n)\mathbf{S}^{\text{H}}{{(n)}}\!=\!\left[ {{{\mathbf{\tilde{z}}}}_{1}}(n),{{{\mathbf{\tilde{z}}}}_{2}}(n),\cdots , {{{\mathbf{\tilde{z}}}}_{M_s}}(n) \right]\in {{\mathbb{C}}^{M_r\times M_s}}$. It can be shown that ${{{\mathbf{\tilde{z}}}}_{l}}(n), l=1,....,M_s,$ are i.i.d. random vectors following distribution $\mathcal{C}\mathcal{N}\left( \mathbf 0, \sigma^2{\mathbf{I}}_{M_r}\right)$.
Let ${{\mathbf{H}}_{j}}=\sqrt{KP\alpha }{{\mathbf{g}}_{j}}\mathbf{h}_{j}^{\text{H}}\mathbf{F}$, ${{\mathbf{x}}_{j}}=\text{vec}\left( {{\mathbf{H}}_{j}} \right)$, $\mathbf{y}(n)=\text{vec}(\mathbf{\tilde{Y}}(n))$, and $\mathbf{z}(n)=\text{vec}(\mathbf{\tilde{Z}}(n))$. The input-output relationship \eqref{eq:tbbeam} can be equivalently written as
\begin{equation}
{\mathbf{y}(n)=\sum\limits_{j=1}^{J}{{{\mathbf{x}}_{j}}}{{c}_{j}}(n)+\mathbf{z}(n)},\label{eq:tbbeam2}
\end{equation}
where ${\mathbf z}(n)\sim \mathcal{CN}(\mathbf 0, \sigma^2\mathbf I_{M_rM_s})$.
Note that \eqref{eq:tbbeam2} is essentially a SIMO MAC, where MMSE-SIC receiver is known to be capacity-achieving~\cite{IEEEhowto:21}. Specifically, the $J$ BD users are ordered according to their channel strength ${{\left\| {{\mathbf{x}}_{j}} \right\|}^{2}}$, based on which the SIC decoding order is determined.
Without loss of generality, assuming that ${{\left\| {{\mathbf{x}}_{1}} \right\|}^{2}}\ge {{\left\| {{\mathbf{x}}_{2}} \right\|}^{2}}\ge \cdots \ge {{\left\| {{\mathbf{x}}_{J}} \right\|}^{2}}$, then the SIC decoding order is $1,2,\cdots ,J$.
Let's focus on BD $j$, where the signals for BDs $1,...,j-1$ have already been decoded and perfectly removed, and those for BDs $j+1,...,J$ are treated as noise. Denote the beamforming vector for BD $j$ as ${{\mathbf{w}}_{j}}\in {{\mathbb{C}}^{M_rM_s\times 1}}$. Then the resulting signal can be written as
\begin{equation}
{{{y}_{j}}(n)=\mathbf{w}_{j}^{\text{H}}{{\mathbf{x}}_{j}}{{c}_{j}}(n)+\mathbf{w}_{j}^{\text{H}}\sum\limits_{i=j+1}^{J}{{{\mathbf{x}}_{i}}}{{c}_{i}}(n)+\mathbf{w}_{j}^{\text{H}}\mathbf{z}(n)}.
\end{equation}
The resulting SINR for BD $j$ is
\begin{equation}
{{{\gamma }_{{{c}_{j}}}}=\frac{{{\left| \mathbf{w}_{j}^{\text{H}}{{\mathbf{x}}_{j}} \right|}^{2}}}{\sum\limits_{i=j+1}^{J}{{{\left| \mathbf{w}_{j}^{\text{H}}{{\mathbf{x}}_{i}} \right|}^{2}}+{{\sigma }^{2}}{{\left\| {{\mathbf{w}}_{j}} \right\|}^{2}}}}}.
\end{equation}
The optimal linear MMSE beamforming that maximizes the SINR is
\begin{equation}
{{{\mathbf{w}}_{j}}={{\Big( \sum\limits_{i=j+1}^{J}{{{\mathbf{x}}_{i}}\mathbf{x}_{i}^{\text{H}}+{{\sigma }^{2}}{{\mathbf{I}}_{M_rM_s}}} \Big)}^{-1}}{{\mathbf{x}}_{j}}},
\end{equation}
and the corresponding maximum SINR is
\begin{equation}
{{{\gamma }_{{{c}_{j}}}}=\mathbf{x}_{j}^{\text{H}}{{\Big( \sum\limits_{i=j+1}^{J}{{{\mathbf{x}}_{i}}\mathbf{x}_{i}^{\text{H}}+{{\sigma }^{2}}{{\mathbf{I}}_{M_rM_s}}} \Big)}^{-1}}{{\mathbf{x}}_{j}}}.\label{BDjmaximumSINR}
\end{equation}
Hence, the sum capacity of the $J$ BDs can be written as
\begin{equation}
\setlength\abovedisplayskip{2pt}
\setlength\belowdisplayskip{2pt}
{{R}_{BD}={\frac{1}{K}}\sum\limits_{j=1}^{J}{\log }_{2} \left( 1+{{\gamma }_{{{c}_{j}}}} \right)},\label{eq:sumrate}
\end{equation}
where the pre-log factor $1/K$ accounts for the fact that in the CSR setup, the symbol duration of the BD users is $K$ times of that of the primary users.

\emph{Theorem 1:} The sum capacity of the BDs in \eqref{eq:sumrate} can be equivalently expressed as
%\begin{small}
\begin{equation}
\setlength\abovedisplayskip{2pt}
\setlength\belowdisplayskip{2pt}
{{{R}_{BD}}=\frac{1}{K}{{\log }_{2}}\Big| {{\mathbf{I}}_{M_rM_s}}+\frac{1}{{{\sigma }^{2}}}\sum\limits_{j=1}^{J}{{{\mathbf{x}}_{j}}\mathbf{x}_{j}^{\text{H}}} \Big|}.\label{eq:sumrate2}
\end{equation}
%\end{small}
\begin{IEEEproof}
Please refer to Appendix A.
\end{IEEEproof}

By subsituting ${\mathbf{x}}_{j}=\text{vec}\left({\mathbf{H}}_{j}\right)=\text{vec}\left( \sqrt{KP\alpha }{{\mathbf{g}}_{j}}\mathbf{h}_{j}^{\text{H}}\mathbf{F} \right)$, the sum capacity of the BDs can be expressed in terms of the precoding matrix $\mathbf{F}$ of the PT, which leads to the following Lemma.

\emph{Lemma 1:} The sum capacity of the BDs ${{R}_{BD}}$ in \eqref{eq:sumrate2} can also be expressed as
\begin{equation}
{{{R}_{BD}}\!=\!\frac{1}{K}{{\log }_{2}}\Big| {{\mathbf{I}}_{M_rM_s}}\!+\!K{\bar P}\alpha\sum\limits_{j=1}^{J}{\left( \left( {{\mathbf{F}}^{\text{H}}}{{\mathbf{h}}_{j}}\mathbf{h}_{j}^{\text{H}}\mathbf{F} \right)\otimes {{\left( {{\mathbf{g}}_{j}}\mathbf{g}_{j}^{\text{H}} \right)}^{\text{T}}} \right)} \Big|}.\label{eq:RBDtrans}
\end{equation}
\begin{IEEEproof}
Please refer to Appendix B.
\end{IEEEproof}
%%>>>>>>>>>>>>>> Section III  >>>>>>>>>>>>>>>>>>>>
\section{Asymptotic Performance Analysis}
To show how the communication rates of symbiotic radio are affected by the number of BDs $J$ and also to explicitly reveal the mutualism relationship between active and passive communications, in this section, we give the asymptotic performance analysis for massive BDs, i.e., as $J$ goes sufficiently large.
We first consider the general MIMO symbiotic radio setup for $M_t\geq 1$, and then consider the special SIMO case with $M_t=1$ to get more insights.
\subsection{MIMO Symbiotic Radio: $M_t\geq 1$}
To obtain tractable asymptotic performance analysis, we assume that the BD channels ${\mathbf{h}}_{j}$ and ${\mathbf{g}}_{j}$ are i.i.d. distributed for different BDs, with $\mathbb{E}\left[ {{\mathbf{h}}_{j}}\mathbf{h}_{j}^{\text{H}} \right]={{\beta }_{h}}{{\mathbf{I}}_{M_t}}$ and $\mathbb{E}\left[ {{\mathbf{g}}_{j}}\mathbf{g}_{j}^{\text{H}} \right]={{\beta }_{g}}{{\mathbf{I}}_{M_r}}$, $\forall j=1,\cdots ,J$, where ${\beta }_{h}$ and ${\beta }_{g}$ are the large-scale channel gains. Under such assumptions, we first study the asymptotic behavior of the capacity of the BDs, for which we have the following Lemma:

\emph{Lemma 2:} For symbiotic radio with massive BDs, i.e., $J\gg  1$, the sum capacity of the BDs in \eqref{eq:RBDtrans} approaches to ${{R}_{BD}}\to \frac{M_r}{K}{{\log }_{2}}\Big| {{\mathbf{I}}_{M_t}}+{JK\bar P\alpha {{\beta }_{h}}{{\beta }_{g}}}\mathbf{F}{{\mathbf{F}}^{\text{H}}} \Big|$.
\begin{IEEEproof}
Please refer to Appendix C.
\end{IEEEproof}
Lemma 2 shows that for symbiotic radio with massive BDs, the sum capacity of the BDs increases monotonically with
the number of BDs $J$, thanks to the multi-user diversity gains.

If the objective is to maximize the asymptotic secondary passive communication rate without considering that of the primary active communication rate, then based on Lemma 2, we have the following optimization problem
\begin{align}
  \max\quad &\frac{{{M}_{r}}}{K}{{\log }_{2}}\Big| {{\mathbf{I}}_{{{M}_{t}}}}+JK\bar{P}\alpha {{\beta }_{h}}{{\beta }_{g}}\mathbf{F}{{\mathbf{F}}^{\text{H}}} \Big|\label{maxasyBDtarget} \\
  {\rm s.t.}\quad &{\rm tr}\left( \mathbf{F}{{\mathbf{F}}^{\text{H}}} \right)=1.\label{maxasyBDconstraint}
\end{align}
Define $\mathbf{Q}\triangleq \mathbf{F}{{\mathbf{F}}^{\text{H}}}$ as the transmit covariance matrix of the PT, and denote the eigenvalues of $\mathbf{Q}$ as ${{\lambda }_{i}}(\mathbf{Q})\ge0$, $i=1,...,M_t$. Then the constraint \eqref{maxasyBDconstraint} can be expressed as
\begin{equation}
{\sum\limits_{i=1}^{{{M}_{t}}}{{{\lambda }_{i}}(\mathbf{Q})}=1.}
\end{equation}
Furthermore, the objective function in \eqref{maxasyBDtarget} can be expressed as
\begin{equation}
\begin{split}
  & \frac{{{M}_{r}}}{K}{{\log }_{2}}\Big| {{\mathbf{I}}_{{{M}_{t}}}}+JK\bar{P}\alpha {{\beta }_{h}}{{\beta }_{g}}\mathbf{F}{{\mathbf{F}}^{\text{H}}} \Big| \\
 =& \frac{{{M}_{r}}}{K}{{\log }_{2}}\Big| {{\mathbf{I}}_{{{M}_{t}}}}+JK\bar{P}\alpha {{\beta }_{h}}{{\beta }_{g}}\mathbf{Q} \Big| \\
 =& \frac{{{M}_{r}}}{K}{{\log }_{2}}\Big( \prod\limits_{i=1}^{{{M}_{t}}}{\left( 1+JK\bar{P}\alpha {{\beta }_{h}}{{\beta }_{g}}{{\lambda }_{i}}\left( \mathbf{Q} \right) \right)} \Big).\label{maxasyBDtargettrans} \\
\end{split}
\end{equation}
According to inequality of arithmetic and geometric mean, we have
\begin{equation}
\begin{split}
& \sqrt[{{M}_{t}}]{\prod\limits_{i=1}^{{{M}_{t}}}{\left( 1+JK\bar{P}\alpha {{\beta }_{h}}{{\beta }_{g}}{{\lambda }_{i}}\left( \mathbf{Q} \right) \right)}}\\
\le & \frac{1}{{{M}_{t}}}\sum\limits_{i=1}^{{{M}_{t}}}{\left( 1+JK\bar{P}\alpha {{\beta }_{h}}{{\beta }_{g}}{{\lambda }_{i}}\left( \mathbf{Q} \right) \right)},
\end{split}
\end{equation}
where the equal sign holds when ${{\lambda }_{1}}(\mathbf{Q})=...={{\lambda }_{M_t}}(\mathbf{Q})=\frac{1}{M_t}$.
As a result, the optimal solution to \eqref{maxasyBDtarget} is achieved when $\mathbf{Q}=\mathbf{F}{{\mathbf{F}}^{\text{H}}}=\frac{1}{{{M}_{t}}}{{\mathbf{I}}_{{{M}_{t}}}}$, and the optimal objective value is
\begin{equation}
 {{R}_{BD} \to R_{BD}^{\mathrm{asym}}=\frac{{{M}_{r}}{{M}_{t}}}{K}{{\log }_{2}}\Big( 1+\frac{JK\bar{P}\alpha {{\beta }_{h}}{{\beta }_{g}}}{{{M}_{t}}}\Big).}
\end{equation}
In this case, the optimal number of data streams ${M}_{s}$ is equal to ${{M}_{t}}$, since $\text{rank}\left( \mathbf{F} \right)={{M}_{t}}$.

Next, we study the asymptotic behavior of the achievable rate of the PT in \eqref{eq:primaryrate}, for which we have the following Lemma:

\emph{Lemma 3:} For symbiotic radio with massive BDs, i.e., $J\gg  1$, the average rate of the PT in \eqref{eq:primaryrate} approaches to ${{R}_{s}}\to {{\log }_{2}}\left| {{\mathbf{I}}_{{{M}_{t}}}}+{\bar P}\mathbf{F}{{\mathbf{F}}^{\text{H}}}\left( \mathbf{H}_{d}^{\text{H}}{{\mathbf{H}}_{d}}+J\alpha {{M}_{r}}{{\beta }_{g}}{{\beta }_{h}}{{\mathbf{I}}_{{{M}_{t}}}} \right) \right|$.
\begin{IEEEproof}
Please refer to Appendix D.
\end{IEEEproof}
If the design objective is to maximize the asymptotic primary active communication rate without considering that of the secondary passive communication rate, then based on Lemma 3, we have the following optimization problem
\begin{small}
\begin{align}
  \max\quad &{{\log }_{2}}\Big| {{\mathbf{I}}_{{{M}_{t}}}}+\bar{P}\mathbf{F}{{\mathbf{F}}^{\text{H}}}\left( \mathbf{H}_{d}^{\text{H}}{{\mathbf{H}}_{d}}+J\alpha {{M}_{r}}{{\beta }_{g}}{{\beta }_{h}}{{\mathbf{I}}_{{{M}_{t}}}} \right) \Big| \label{maxasympriobj}\\
  \text{s.t.}\quad &\text{tr}\left( \mathbf{F}{{\mathbf{F}}^{\text{H}}} \right)=1.
\end{align}
\end{small}
Define ${{\mathbf{\bar{H}}}_{eq}}$ as the equivalent MIMO channel matrix in \eqref{maxasympriobj}, where $\mathbf{\bar{H}}_{eq}^{\text{H}}{{\mathbf{\bar{H}}}_{eq}}\triangleq \mathbf{H}_{d}^{\text{H}}{{\mathbf{H}}_{d}}+J\alpha {{M}_{r}}{{\beta }_{g}}{{\beta }_{h}}{{\mathbf{I}}_{{{M}_{t}}}}$.
Let the singular value decomposition (SVD) of the matrix direct-link ${{\mathbf{H}}_{d}}$ as ${{\mathbf{H}}_{d}}={{\mathbf{U}}_{{{{\mathbf{H}}}_{d}}}}{{\mathbf{\Sigma }}_{{{{\mathbf{H}}}_{d}}}}\mathbf{V}_{{{{\mathbf{H}}}_{d}}}^{\text{H}}$, with $\mathbf{\Sigma }_{{{{\mathbf{H}}}_{d}}}^{\text{H}}{{\mathbf{\Sigma }}_{{{{\mathbf{H}}}_{d}}}}$ a diagonal matrix containing the eigenvalues of $\mathbf{H}_{d}^{\text{H}}{{\mathbf{H}}_{d}}$. ${{\mathbf{U}}_{{{{\mathbf{H}}}_{d}}}}\in {{\mathbb{C}}^{{{M}_{r}}\times {{M}_{r}}}}$ and ${{\mathbf{V}}_{{{{\mathbf{H}}}_{d}}}}\in {{\mathbb{C}}^{{{M}_{t}}\times {{M}_{t}}}}$ are both unitary matrices.
$\mathbf{\bar{H}}_{eq}^{\text{H}}{{\mathbf{\bar{H}}}_{eq}}$ can be accordingly expressed as
\begin{equation}
\begin{split}
   \mathbf{\bar{H}}_{eq}^{\text{H}}{{{\mathbf{\bar{H}}}}_{eq}}& ={{\left( {{\mathbf{U}}_{{{\mathbf{H}}_{d}}}}{{\mathbf{\Sigma }}_{{{\mathbf{H}}_{d}}}}\mathbf{V}_{{{\mathbf{H}}_{d}}}^{\text{H}} \right)}^{\text{H}}}\left( {{\mathbf{U}}_{{{\mathbf{H}}_{d}}}}{{\mathbf{\Sigma }}_{{{\mathbf{H}}_{d}}}}\mathbf{V}_{{{\mathbf{H}}_{d}}}^{\text{H}} \right)\\
  & +J\alpha {{M}_{r}}{{\beta }_{g}}{{\beta }_{h}}{{\mathbf{V}}_{{{\mathbf{H}}_{d}}}}\mathbf{V}_{{{\mathbf{H}}_{d}}}^{\text{H}} \\
 & ={{\mathbf{V}}_{{{\mathbf{H}}_{d}}}}\left( \mathbf{\Sigma }_{{{\mathbf{H}}_{d}}}^{\text{H}}{{\mathbf{\Sigma }}_{{{\mathbf{H}}_{d}}}}+J\alpha {{M}_{r}}{{\beta }_{g}}{{\beta }_{h}}{{\mathbf{I}}_{{{M}_{t}}}} \right)\mathbf{V}_{{{\mathbf{H}}_{d}}}^{\text{H}}. \\
\end{split}
\end{equation}
According to \cite{IEEEhowto:30}, the optimum precoder that maximizes the primary asymptotic rate in \eqref{maxasympriobj} is given by
\begin{equation}
{\mathbf{F}=\frac{1}{\sqrt{{{M}_{t}}}}{{\mathbf{V}}_{{{\mathbf{H}}_{d}}}}{{\mathbf{P}}^{\frac{1}{2}}}},
\end{equation}
%where ${{\mathbf{V}}_{{{\mathbf{H}}_{d}}}}$ can be obtained by invoking the (reduced) singular value decomposition (SVD) of the channel matrix, ${{\mathbf{H}}_{d}}={{\mathbf{U}}_{{{{\mathbf{H}}}_{d}}}}{{\mathbf{\Sigma }}_{{{{\mathbf{H}}}_{d}}}}\mathbf{V}_{{{{\mathbf{H}}}_{d}}}^{\text{H}}$ with $\mathbf{\Sigma }_{{{{\mathbf{H}}}_{d}}}^{\text{H}}{{\mathbf{\Sigma }}_{{{{\mathbf{H}}}_{d}}}}$ a diagonal matrix containing the positive eigenvalues of $\mathbf{H}_{d}^{\text{H}}{{\mathbf{H}}_{d}}$. ${{\mathbf{U}}_{{{{\mathbf{H}}}_{d}}}}\in {{\mathbb{C}}^{{{M}_{r}}\times \text{rank}\left( {{{\mathbf{H}}}_{d}} \right)}}$ and ${{\mathbf{V}}_{{{{\mathbf{H}}}_{d}}}}\in {{\mathbb{C}}^{{{M}_{t}}\times \text{rank}\left( {{{\mathbf{H}}}_{d}} \right)}}$ are both semi-unitary matrixes.
where $\mathbf{P}$ is the diagonal power allocation matrix given by
\begin{equation}
{\mathbf{P}=\text{diag}\left( P_{0}^{*},\cdots ,P_{{{M}_{s}}-1}^{*} \right)},
\end{equation}
where ${M}_{s}={{M}_{t}}$ since the matrix $\mathbf{\bar{H}}_{eq}^{\text{H}}{{\mathbf{\bar{H}}}_{eq}}$ is full rank, and the optimum $P_{0}^{*},\cdots ,P_{{{M}_{s}}-1}^{*}$ is given by the waterfilling policy that can be obtained as the fixed point of the equations
\begin{equation}
{P_{i}^{*}=\frac{1-\text{MMS}{{\text{E}}_{i}}\left( P_{i}^{*} \right)}{\frac{1}{{{M}_{s}}}\sum\limits_{q=0}^{{{M}_{s}}-1}{\left( 1-\text{MMS}{{\text{E}}_{q}}\left( P_{q}^{*} \right) \right)}},i=0,\cdots ,{{M}_{s}}-1,}
\end{equation}
where
\begin{equation}
{\text{MMS}{{\text{E}}_{i}}\left( P_{i}^{*} \right)=\frac{1}{1+\frac{{\bar{P}}}{{{M}_{t}}}P_{i}^{*}\left({{\lambda }_{i}}\left( \mathbf{H}_{d}^{\text{H}}{{\mathbf{H}}_{d}}\right)+J\alpha {{M}_{r}}{{\beta }_{g}}{{\beta }_{h}} \right)}.}
\end{equation}
Then the resulting primary asymptotic rate is
\begin{equation}
\begin{split}
{R}_{s}& \to R_{s}^{\mathrm{asym}}\\
& =\sum\limits_{i=0}^{{{M}_{t}}-1}{{{\log }_{2}}\Big( 1+\frac{\bar{P}}{{M}_{t}}P_{i}^{*}\left({{\lambda }_{i}}\left( \mathbf{H}_{d}^{\text{H}}{{\mathbf{H}}_{d}}\right)+J\alpha {{M}_{r}}{{\beta }_{g}}{{\beta }_{h}} \right) \Big)}\label{asymprirate}.
\end{split}
\end{equation}
It is observed from \eqref{asymprirate} that the asymptotic average primary communication rate also increases with the number of BDs $J$, thanks to the multi-path diversity created to the primary communication link via passive backscattering. To more explicitly show the mutualism relationship between active and passive communications, we consider the special case of SIMO setup to gain more insights, as revealed in the following subsection.

\subsection{SIMO Symbiotic Radio: $M_t=1$}
When the transmitter has only one antenna, i.e., $M_t=1$, the transmit precoding matrix $\mathbf{F}$ degenerates to a scalar $f$, with ${{\left| f \right|}^{2}}=1$, and the MISO channel vector $\mathbf{h}_{j}$ from the PT to BD $j$ degenerates to a channel coefficient ${h}_{j}$. In this case, Lemma 2 degenerates to the following corollary.

\emph{Corollary:} For SIMO symbiotic radio with massive BDs, i.e., $J\gg  1$, the sum capacity of the BDs in \eqref{eq:RBDtrans} approaches to
\begin{equation}
{R_{BD}\to {R}_{BD}^{\mathrm{asym}}=\frac{M_r}{K}{\log }_{2} \left( 1+{JK\bar P\alpha {{\beta }_{h}}{{\beta }_{g}}} \right).}\label{eq:RBDJ>>1L=1}
\end{equation}
Similarly, the asymptotic average rate of the PT in Lemma 3 for SIMO setup degenerates to the following result:

\emph{Corollary:} For SIMO symbiotic radio with massive BDs, i.e., $J\gg  1$, $R_{s}$ in Lemma 3 can be written as
\begin{equation}
{{{R}_{s}}\to {R}_{s}^{\mathrm{asym}}={{\log }_{2}}\left( 1+{\bar P\left( {{\left\| {{\mathbf{h}}_{d}} \right\|}^{2}}+J\alpha {{M}_{r}}{{\beta }_{g}}{{\beta }_{h}} \right)} \right)}\label{eq:RsJ>>1L=1}.
\end{equation}
It is not difficult to see that ${R}_{s}$ in both MIMO symbiotic radio setup and SIMO symbiotic radio setup in Lemma 3 and \eqref{eq:RsJ>>1L=1} respectively
increases monotonically with the number of BDs $J$. Thus, with more BDs connected to the symbiotic radio system, the enhancement to the primary transmission becomes more significant. Based on \eqref{eq:RBDJ>>1L=1} and \eqref{eq:RsJ>>1L=1}, by eliminating the common variables $J\bar P\alpha\beta_h\beta_g$, we have the following result:

\emph{Theorem 2:} For SIMO symbiotic radio with massive BDs, i.e., $J\gg 1$, the asymptotic primary communication rate ${R}_{s}^{\mathrm{asym}}$ can be expressed in closed-form in terms of the asymptotic secondary sum capacity ${R}_{BD}^{\mathrm{asym}}$ as:
\begin{equation}
{R_{s}^{\mathrm{asym}}={{\log }_{2}}\bigg( 1+\bar{P}{{\left\| {{\mathbf{h}}_{d}} \right\|}^{2}}+\frac{{{M}_{r}}}{K}\left( {{2}^{\frac{K}{{{M}_{r}}}R_{BD}^{\mathrm{asym}}}}-1 \right) \bigg)}.\label{eq:RSRBD}
\end{equation}
It is evident that ${R}_{s}^{\mathrm{asym}}$ in \eqref{eq:RSRBD} monotonically increases with ${R}_{BD}^{\mathrm{asym}}$, which clearly reveals the mutualism relationship of symbiotic radio with massive BDs.
Furthermore, when the direct link from the PT to AP ${{\mathbf{h}}_{d}}$ is a line of
sight (LoS) link, then ${{\left\| {{\mathbf{h}}_{d}} \right\|}^{2}}$ is proportional to the number of receive antennas $M_r$ at the AP, i.e., ${{\left\| {{\mathbf{h}}_{d}} \right\|}^{2}}= {{\beta }_{{{h}_{d}}}}{{M}_{r}}$ for some ${{\beta }_{{{h}_{d}}}}$. Then \eqref{eq:RSRBD} can be expressed as
\begin{equation}
{R_{s}^{\mathrm{asym}}={{\log }_{2}}\left( 1+\bar{P}{{\beta }_{{{h}_{d}}}}{{M}_{r}}+\frac{{{M}_{r}}}{K}\left( {{2}^{\frac{K}{{{M}_{r}}}R_{BD}^{\mathrm{asym}}}}-1 \right) \right).}\label{eq:RSRBDtrans}
\end{equation}
\begin{figure}[!t]
  \centering
  \centerline{\includegraphics[height=3in, width=3.5in]{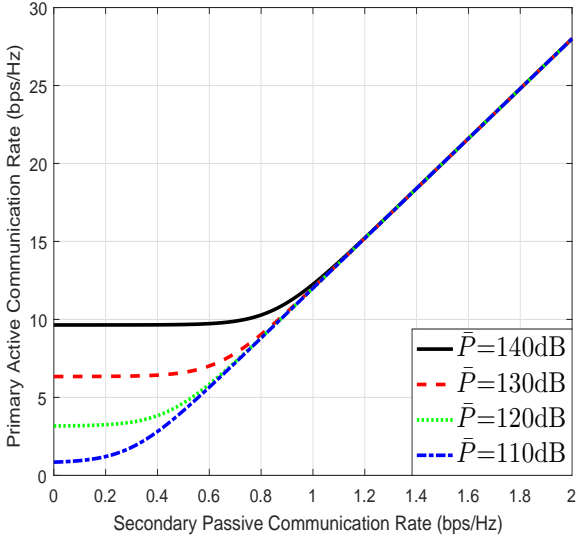}}
  \caption{The asymptotic primary active communication rate versus secondary passive communication rate in SIMO symbiotic radio.}
  \label{system model}
   \vspace{-0.5cm}
  \end{figure}

Fig. 2 gives an example plot of \eqref{eq:RSRBDtrans} with ${{\beta }_{{{h}_{d}}}}=-120$ dB, $M_r=8$ and $K=128$, for four different transmit SNR values $\bar P$. It is worth mentioning that while \eqref{eq:RSRBDtrans} was derived for asymptotic setup with $J\gg 1$, it is also applicable for the extreme case with $J=0$ or ${R}_{BD}^{\mathrm{asym}}=0$, for which case the third term inside the logarithm of \eqref{eq:RSRBDtrans} vanishes. Therefore, Fig. 2 plots ${R}_{s}^{\mathrm{asym}}$ versus ${R}_{BD}^{\mathrm{asym}}$ in \eqref{eq:RSRBDtrans}, starting from ${R}_{BD}^{\mathrm{asym}}=0$. It is observed from Fig. 2 that when the secondary passive communication rate is small, i.e., when $J$ is small and/or the cascaded channel $\sqrt{\alpha}{{h}_{j}}{\mathbf{g}_j}$ is weak, increasing ${R}_{BD}^{\mathrm{asym}}$ has a negligible impact on the primary active communication rate. However, once ${R}_{BD}^{\mathrm{asym}}$ exceeds certain threshold, ${R}_{s}^{\mathrm{asym}}$ increases almost linearly with ${R}_{BD}^{\mathrm{asym}}$. This thus demonstrates that the mutualism of symbiotic radio can only be fully exploited for sufficiently large BDs. It is also interesting to observe that as ${R}_{BD}^{\mathrm{asym}}$  becomes sufficiently large, ${R}_{s}^{\mathrm{asym}}$  for different $\bar P$ values merge. This can be verified by taking the derivative of ${R}_{s}^{\mathrm{asym}}$ in \eqref{eq:RSRBDtrans} with respect to $\bar P$, which vanishes as ${R}_{BD}^{\mathrm{asym}}$ gets sufficiently large.

\section{Precoding Optimization}
In this section, for any given finite number of BDs $J$ where the asymptotic results in the previous section no longer hold, we consider the transmit precoding optimization problem to maximize the average primary communication rate in
\eqref{eq:primaryrate}, under the sum rate constraint of the BDs in \eqref{eq:RBDtrans}. The following problem can be formulated:
\begin{small}
\begin{align}
 \notag (\text{P}1)\quad\quad &\\
  \underset{M_s\leq M_t}{\underset{\mathbf{F}\in \mathbb{C}^{M_t \times M_s},}{\mathop{\max }}}\,& {{\mathbb{E}}_{\mathbf{c}(n)}}\left[ {{\log }_{2}}\left| {{\mathbf{I}}_{M_r}}+{\bar P}{{\mathbf{H}}_{eq}}\left( \mathbf{c}\left( n \right) \right)\mathbf{F}{{\mathbf{F}}^{\text{H}}}\mathbf{H}_{eq}^{\text{H}}\left( \mathbf{c}\left( n \right) \right) \right| \right] \\
 \notag {\rm s.t.}\quad\frac{1}{K}&{{\log }_{2}}\Big| {{\mathbf{I}}_{M_rM_s}}+{K\bar P\alpha }\sum\limits_{j=1}^{J}{\Big( \big( {{\mathbf{F}}^{\text{H}}}{{\mathbf{h}}_{j}}\mathbf{h}_{j}^{\text{H}}\mathbf{F} \big)\otimes {{\big( {{\mathbf{g}}_{j}}\mathbf{g}_{j}^{\text{H}} \big)}^{\text{T}}} \Big)} \Big|\\
 &  \quad \quad \quad \quad \quad \quad \quad \quad \quad \quad \quad \quad \quad \quad \quad \quad \ge {{r}_{BD}}, \label{eq:constraintBDs}\\
 {\rm tr}& \big( \mathbf{F}{{\mathbf{F}}^{\text{H}}} \big)=1,
\end{align}
\end{small}
where ${r}_{BD}$ denotes the given targeting threshold for the sum rate of the $J$ BDs.
Note that the optimization variables in (P1) include the number of data streams $M_s$, as well as the precoding matrix $\mathbf{F}$. Problem (P1) is non-convex, which is difficult to be directly solved. Fortunately, it can be transformed to an equivalent convex problem of optimizing the transmit covariance matrix. To this end, define $\mathbf{Q}=\mathbf{F}{{\mathbf{F}}^{\text{H}}}$, which is the transmit covariance matrix of the PT. Then problem (P1) can be equivalently written as
\begin{small}
\begin{align}
  \notag (\text{P}1.1)\quad\quad &\\
  \underset{M_s\leq M_t}{\underset{\mathbf{F}\in \mathbb{C}^{M_t \times M_s},\mathbf{Q},}{\mathop{\max }}}\,& {{\mathbb{E}}_{\mathbf{c}(n)}}\left[ {{\log }_{2}}\left| {{\mathbf{I}}_{M_r}}+{\bar P}{{\mathbf{H}}_{eq}}\left( \mathbf{c}\left( n \right) \right)\mathbf{Q}\mathbf{H}_{eq}^{\text{H}}\left( \mathbf{c}\left( n \right) \right) \right| \right] \label{eq:primaryFW}\\
 \notag {\rm s.t.}\quad\frac{1}{K}&{{\log }_{2}}\Big| {{\mathbf{I}}_{M_rM_s}}+{K\bar P\alpha }\sum\limits_{j=1}^{J}{\Big( \big( {{\mathbf{F}}^{\text{H}}}{{\mathbf{h}}_{j}}\mathbf{h}_{j}^{\text{H}}\mathbf{F} \big)\otimes {{\big( {{\mathbf{g}}_{j}}\mathbf{g}_{j}^{\text{H}} \big)}^{\text{T}}} \Big)} \Big|\\
 &  \quad \quad \quad \quad \quad \quad \quad \quad \quad \quad \quad \quad \quad \quad \quad \ge {{r}_{BD}}, \label{RBDcomplex}\\
  {\rm tr}& (\mathbf{Q})=1,\\
 \mathbf{Q}& =\mathbf{F}{{\mathbf{F}}^{\text{H}}}\label{Q=FF^H}.
\end{align}
\end{small}
(P1.1) is still non-convex since the constraint \eqref{RBDcomplex} and \eqref{Q=FF^H} are non-convex. In the following, to find an equivalent convex transformation of (P1.1), we first consider the special case of one single BD, i.e., $J=1$, and then consider the more general case with $J\geq 1$.
\subsection{Single BD: $J=1$}
For the special case of single BD with $J = 1$, the left hand side of \eqref{RBDcomplex} can be simplified as:
\begin{equation}
\begin{split}
  {{R}_{1}}& =\frac{1}{K}{{\log }_{2}}\left| {{\mathbf{I}}_{M_rM_s}}+{K\bar P\alpha }\left( {{\mathbf{F}}^{\text{H}}}{{\mathbf{h}}_{1}}\mathbf{h}_{1}^{\text{H}}\mathbf{F} \right)\otimes {{\left( {{\mathbf{g}}_{1}}\mathbf{g}_{1}^{\text{H}} \right)}^{\text{T}}} \right| \\
 & =\frac{1}{K}{{\log }_{2}}\left| {{\mathbf{I}}_{M_rM_s}}+{K\bar P\alpha }\left( {{\mathbf{F}}^{\text{H}}}{{\mathbf{h}}_{1}}\otimes \mathbf{g}_{1}^{*} \right)\left( \mathbf{h}_{1}^{\text{H}}\mathbf{F}\otimes \mathbf{g}_{1}^{\text{T}} \right) \right| \\
 & =\frac{1}{K}{{\log }_{2}}\left( 1+{K\bar P\alpha }\left( \mathbf{h}_{1}^{\text{H}}\mathbf{F}\otimes \mathbf{g}_{1}^{\text{T}} \right)\left( {{\mathbf{F}}^{\text{H}}}{{\mathbf{h}}_{1}}\otimes \mathbf{g}_{1}^{*} \right) \right) \\
 & =\frac{1}{K}{{\log }_{2}}\left( 1+{K\bar P\alpha {{\left\| {{\mathbf{g}}_{1}} \right\|}^{2}}\mathbf{h}_{1}^{\text{H}}\mathbf{F}{{\mathbf{F}}^{\text{H}}}{{\mathbf{h}}_{1}}} \right),\label{J=1RBDF}
\end{split}
\end{equation}
where the second and last equalities follow from the identity $\left( \mathbf{A}\otimes \mathbf{B} \right)\left( \mathbf{C}\otimes \mathbf{D} \right)=\mathbf{AC}\otimes \mathbf{BD}$, and the third equality follows from the Weinstein-Aronszajn identity $\left| {{\mathbf{I}}_{m}}+\mathbf{AB} \right|=\left| {{\mathbf{I}}_{n}}+\mathbf{BA} \right|$.
By substituting \eqref{J=1RBDF} and $\mathbf{Q}=\mathbf{F}{{\mathbf{F}}^{\text{H}}}$ into the left hand side of \eqref{RBDcomplex}, (P1.1) can be transformed to:
\begin{align}
 \notag (\text{P}2)\quad & \\
 \underset{\mathbf{Q}}{\mathop{\max }}\quad & {{\mathbb{E}}_{\mathbf{c}(n)}}\left[ {{\log }_{2}}\left| {{\mathbf{I}}_{M_r}}+{\bar P}{{\mathbf{H}}_{eq}}\left( \mathbf{c}\left( n \right) \right)\mathbf{QH}_{eq}^{\text{H}}\left( \mathbf{c}\left( n \right) \right) \right| \right]\label{J=1Rsexpectation}\\
 {\rm s.t.}\quad &\frac{1}{K}{{\log }_{2}}\left( 1+{K\bar P\alpha {{\left\| {{\mathbf{g}}_{1}} \right\|}^{2}}\mathbf{h}_{1}^{\text{H}}\mathbf{Q}{{\mathbf{h}}_{1}}} \right)\ge {{r}_{BD}},\label{J=1RBDQ}\\
 & {\rm tr}(\mathbf{Q})=1,\mathbf{Q}\succeq0\label{trace=1}.
\end{align}
Note that in (P2), the only optimization variable is the transmit covariance matrix $\mathbf{Q}$, and we have temporarily removed the constraints $\mathbf{Q}=\mathbf{F}{{\mathbf{F}}^{\text{H}}}$, $\mathbf{F}\in \mathbb{C}^{M_t\times M_s}$, and $M_s \leq M_t$. This does not affect the equivalence between P(1.1) and (P2), since once the optimal transmit covariance matrix $\mathbf{Q}^\star$ to (P2) is obtained, we can always find the corresponding precoding matrix $\mathbf{F}$ and the number of data streams $M_s$ satisfying the constraint of (P1.1). One straightforward method is by applying eigenvalue decomposition to $\mathbf{Q}^\star$. Specifically, let the (reduced) EVD of $\mathbf{Q}^\star$ be $\mathbf{Q}^\star=\mathbf{U\Sigma }{{\mathbf{U}}^{\text{H}}}$, where $\mathbf{U}\in {{\mathbb{C}}^{{{M}_{t}}\times r}}$ is a semi-unitary matrix and $\mathbf{\Sigma}$ is an $r\times r$ diagonal matrix whose diagonal elements are the positive eigenvalues of $\mathbf{Q}^\star$, and $r\leq M_t$ is the rank of $\mathbf{Q}^\star$.Then, to find $M_s$ and $\mathbf{F}$ that are feasible to (P1.1), we may simply let $M_s=r$, and ${\mathbf{F}}=\mathbf{U}\mathbf{\Sigma }^{1/2}$.
Obviously, the corresponding solution is optimal to (P1.1) since it achieves the same optimal value as its relaxed problem (P2). Therefore, the remaining task is to solve (P2) by optimizing the transmit covariance matrix $\mathbf{Q}$. It is not difficult to see that (P2) is a convex optimization problem, which in principle can be solved by standard convex optimization techniques. However, the difficulty lies in that the objective function of (P2) involves an expectation with respect to the random BD symbols $\mathbf{c}(n)$. To tackle this issue, we will approximate \eqref{J=1Rsexpectation} by its sample average or upper bound, and convert the problem into a more tractable convex optimization problem. The details will be presented in Subsection IV B, since similar techniques will also be used for the general case with multiple BDs presented in the next subsection.
\subsection{Multiple BDs: $J\geq 1$}
In this subsection, we consider the general case of (P1.1) with multiple BDs.
In this case, by using Theorem 1, the left hand side of \eqref{RBDcomplex} can be equivalently expressed as \eqref{eq:sumrate2}. Define $\mathbf{X}=\left[ {{\mathbf{x}}_{1}},{{\mathbf{x}}_{2}},\cdots ,{{\mathbf{x}}_{J}} \right]$, we have
\begin{equation}
\begin{split}
 {{R}_{BD}}& =\frac{1}{K}{{\log }_{2}}\Big| {{\mathbf{I}}_{M_rM_s}}+\frac{1}{{{\sigma }^{2}}}\sum\limits_{j=1}^{J}{{{\mathbf{x}}_{j}}\mathbf{x}_{j}^{\text{H}}} \Big| \\
 & =\frac{1}{K}{{\log }_{2}}\Big| {{\mathbf{I}}_{M_rM_s}}+\frac{1}{{{\sigma }^{2}}}\mathbf{X}{{\mathbf{X}}^{\text{H}}} \Big| \\
 & =\frac{1}{K}{{\log }_{2}}\Big| {{\mathbf{I}}_{J}}+\frac{1}{{{\sigma }^{2}}}{{\mathbf{X}}^{\text{H}}}\mathbf{X} \Big| \\
 & =\frac{1}{K}{{\log }_{2}}\Big| {{\mathbf{I}}_{J}}+\frac{1}{{{\sigma }^{2}}}{{\mathbf{X}}^{\text{T}}}{{\mathbf{X}}^{*}} \Big|\label{eq:multiBDratetransform},
\end{split}
\end{equation}
where the second last equality follows from the Weinstein-Aronszajn identity $\left| {{\mathbf{I}}_{m}}+\mathbf{AB} \right|=\left| {{\mathbf{I}}_{n}}+\mathbf{BA} \right|$, and the last equality follows from the identity $\left| {{\mathbf{A}}^{\text{T}}} \right|=\left| \mathbf{A} \right|$.
By using the identity $\text{vec}({{\mathbf{A}}_{1}}{{\mathbf{A}}_{2}}{{\mathbf{A}}_{3}})=\left( \mathbf{A}_{3}^{\text{T}}\otimes {{\mathbf{A}}_{1}} \right)\text{vec}({{\mathbf{A}}_{2}})$, we have
\begin{equation}
\begin{split}
  {{\mathbf{x}}_{j}}& =\text{vec}\left( \sqrt{KP\alpha }{{\mathbf{g}}_{j}}\mathbf{h}_{j}^{\text{H}}\mathbf{F} \right) \\
 & =\sqrt{KP\alpha }\left( {{\mathbf{F}}^{\text{T}}}\otimes {{\mathbf{g}}_{j}}\mathbf{h}_{j}^{\text{H}} \right)\text{vec}\left( {{\mathbf{I}}_{M_t}} \right).\label{eq:xj}
\end{split}
\end{equation}
Then $\mathbf{X}$ can be expressed as \eqref{eq:X} shown at the top of the next page.
\newcounter{TempEqCnt} % 创建临时变量TempEqCnt
\setcounter{TempEqCnt}{\value{equation}} % 将当前公式序号 赋给TempEqCnt
\setcounter{equation}{45} % 当前公式序号变为x，x等于长公式应有的序号减1.

\begin{figure*}[ht] %hb代表放在文章底部，%ht为放在文章顶部
%\begin{small}
\begin{equation}
\begin{split}
  \mathbf{X}& =\sqrt{KP\alpha }\left[ \left( {{\mathbf{F}}^{\text{T}}}\otimes {{\mathbf{g}}_{1}}\mathbf{h}_{1}^{\text{H}} \right)\text{vec}\left( {{\mathbf{I}}_{M_t}} \right),\left( {{\mathbf{F}}^{\text{T}}}\otimes {{\mathbf{g}}_{2}}\mathbf{h}_{2}^{\text{H}} \right)\text{vec}\left( {{\mathbf{I}}_{M_t}} \right),\cdots ,\left( {{\mathbf{F}}^{\text{T}}}\otimes {{\mathbf{g}}_{J}}\mathbf{h}_{J}^{\text{H}} \right)\text{vec}\left( {{\mathbf{I}}_{M_t}} \right) \right] \\
 & =\sqrt{KP\alpha }\left[ \left( {{\mathbf{F}}^{\text{T}}}\otimes {{\mathbf{g}}_{1}}\mathbf{h}_{1}^{\text{H}} \right),\left( {{\mathbf{F}}^{\text{T}}}\otimes {{\mathbf{g}}_{2}}\mathbf{h}_{2}^{\text{H}} \right),\cdots ,\left( {{\mathbf{F}}^{\text{T}}}\otimes {{\mathbf{g}}_{J}}\mathbf{h}_{J}^{\text{H}} \right) \right]\left[ \begin{matrix}
   \text{vec}\left( {{\mathbf{I}}_{M_t}} \right) & {\mathbf{0}} & \cdots  & {\mathbf{0}}  \\
   {\mathbf{0}} & \text{vec}\left( {{\mathbf{I}}_{M_t}} \right) & \cdots  & {\mathbf{0}}  \\
   \vdots  & \vdots & \ddots  & \vdots  \\
   {\mathbf{0}} & {\mathbf{0}} & \cdots  & \text{vec}\left( {{\mathbf{I}}_{M_t}} \right)  \\
\end{matrix} \right] \\
 & =\sqrt{KP\alpha }\left( {{\mathbf{F}}^{\text{T}}}\otimes \left[ {{\mathbf{g}}_{1}}\mathbf{h}_{1}^{\text{H}},{{\mathbf{g}}_{2}}\mathbf{h}_{2}^{\text{H}},\cdots ,{{\mathbf{g}}_{J}}\mathbf{h}_{J}^{\text{H}} \right] \right)\left[ \begin{matrix}
   \text{vec}\left( {{\mathbf{I}}_{M_t}} \right) & {\mathbf{0}} & \cdots  & {\mathbf{0}}  \\
   {\mathbf{0}} & \text{vec}\left( {{\mathbf{I}}_{M_t}} \right) & \cdots  & {\mathbf{0}}  \\
   \vdots  & \vdots & \ddots  & \vdots  \\
   {\mathbf{0}} & {\mathbf{0}} & \cdots  & \text{vec}\left( {{\mathbf{I}}_{M_t}} \right)  \\
\end{matrix} \right]\label{eq:X}
\end{split}
\end{equation}
 \hrulefill
\end{figure*}
For notational convenience, we let $\mathbf{H}=\left[ {{\mathbf{g}}_{1}}\mathbf{h}_{1}^{\text{H}},{{\mathbf{g}}_{2}}\mathbf{h}_{2}^{\text{H}},\cdots ,{{\mathbf{g}}_{J}}\mathbf{h}_{J}^{\text{H}} \right]\in {{\mathbb{C}}^{M_r\times M_tJ}}$ and
\begin{equation}
\mathbf{\Psi} =\left[ \begin{matrix}
   \text{vec}\left( {{\mathbf{I}}_{M_t}} \right) & {\mathbf{0}} & \cdots  & {\mathbf{0}}  \\
   {\mathbf{0}} & \text{vec}\left( {{\mathbf{I}}_{M_t}} \right) & \cdots  & {\mathbf{0}}  \\
   \vdots  & \vdots & \ddots  & \vdots  \\
   {\mathbf{0}} & {\mathbf{0}} & \cdots  & \text{vec}\left( {{\mathbf{I}}_{M_t}} \right)  \\
\end{matrix} \right]\in {{\mathbb{R}}^{M_t^{2}J\times J}},
\end{equation}
so that $\mathbf{X}=\sqrt{KP\alpha }\left( {{\mathbf{F}}^{\text{T}}}\otimes \mathbf{H} \right)\mathbf{\Psi }$. Then $R_{BD}$ in \eqref{eq:multiBDratetransform} can be expressed as
\begin{equation}
\begin{split}
 {{R}_{BD}}& =\frac{1}{K}{{\log }_{2}}\Big| {{\mathbf{I}}_{J}}+\frac{1}{{{\sigma }^{2}}}{{\mathbf{X}}^{\text{T}}}{{\mathbf{X}}^{*}} \Big| \\
 & =\frac{1}{K}{{\log }_{2}}\Big| {{\mathbf{I}}_{J}}+{K\bar P\alpha }{{\mathbf{\Psi }}^{\text{T}}}{{\left( {{\mathbf{F}}^{\text{T}}}\otimes \mathbf{H} \right)}^{\text{T}}}{{\left( {{\mathbf{F}}^{\text{T}}}\otimes \mathbf{H} \right)}^{*}}{{\mathbf{\Psi }}^{*}} \Big| \\
 & =\frac{1}{K}{{\log }_{2}}\Big| {{\mathbf{I}}_{J}}+{K\bar P\alpha }{{\mathbf{\Psi }}^{\text{H}}}\left( \mathbf{F}{{\mathbf{F}}^{\text{H}}}\otimes {{\left( {{\mathbf{H}}^{\text{H}}}\mathbf{H} \right)}^{\text{T}}} \right)\mathbf{\Psi } \Big|, \\\label{eq:RBDnew}
\end{split}
\end{equation}
where the last equality follows from the identities ${{\left( \mathbf{A}\otimes \mathbf{B} \right)}^{\text{T}}}={{\mathbf{A}}^{\text{T}}}\otimes {{\mathbf{B}}^{\text{T}}}$, ${{\left( \mathbf{A}\otimes \mathbf{B} \right)}^{*}}={{\mathbf{A}}^{*}}\otimes {{\mathbf{B}}^{*}}$, $\left( \mathbf{A}\otimes \mathbf{B} \right)\left( \mathbf{C}\otimes \mathbf{D} \right)=\mathbf{AC}\otimes \mathbf{BD}$ and $\mathbf{\Psi}\in {\mathbb{R}}$. When $J=1$, $\mathbf{H}={{\mathbf{g}}_{1}}\mathbf{h}_{1}^{\text{H}}$, $\mathbf{\Psi }=\text{vec}\left( {{\mathbf{I}}_{{{M}_{t}}}} \right)$, \eqref{eq:RBDnew} can be simplified to \eqref{J=1RBDF}.
By replacing $R_{BD}$ in \eqref{RBDcomplex} with \eqref{eq:RBDnew} and letting $\mathbf{Q}=\mathbf{F}{{\mathbf{F}}^{\text{H}}}$, we transform (P1.1) into
\begin{align}
 \notag(\text{P}3)\quad &\\
 \underset{\mathbf{Q}}{\mathop{\max }}\quad &{{\mathbb{E}}_{\mathbf{c}(n)}}\left[ {{\log }_{2}}\left| {{\mathbf{I}}_{M_r}}+{\bar P}{{\mathbf{H}}_{eq}}\left( \mathbf{c}\left( n \right) \right)\mathbf{QH}_{eq}^{\text{H}}\left( \mathbf{c}\left( n \right) \right) \right| \right]\label{eq:generalRs} \\
 {\rm s.t.}\quad &\frac{1}{K}{{\log }_{2}}\Big| {{\mathbf{I}}_{J}}+{K\bar P\alpha }{{\mathbf{\Psi }}^{\text{H}}}\Big( \mathbf{Q}\otimes {{\left( {{\mathbf{H}}^{\text{H}}}\mathbf{H} \right)}^{\text{T}}} \Big)\mathbf{\Psi } \Big|\ge {{r}_{BD}},\label{eq:RBDwithQ} \\
 & {\rm tr}(\mathbf{Q})=1,\mathbf{Q}\succeq0.
\end{align}
After solving (P3), we can obtain $\mathbf{F}$ by using the same way as discussed in Subsection V A.

Note that similar to (P2), the transmit covariance matrix $\mathbf{Q}$ in (P3) is optimized to maximize the expected primary communication rate, with the expectation taken with respect to the random BD symbols $\mathbf c(n)$. In the following, we propose two solution approaches to (P3), termed \emph{sample-average based solution} and \emph{upper bound based solution}, which have different trade-offs on performance and computational complexity.
%which are applicable to scenarios with different primary transmission rate requirements and have different computational complexity.
\subsubsection{Sample-Average Based Solution}
For the sample-average based solution, the expectation of the primary communication rate in \eqref{eq:generalRs} is approximated by its sample average. Specifically, let $\mathbf c_s, s=1,...,S$ be $S$ independent realizations of $\mathbf c(n)$ following its distribution $f(\mathbf{c}(n))$. Then when $S$ is sufficiently large, based on the law of large numbers, (P3) can be approximated as
\begin{align}
 \notag(\text{P}4)\quad &\\
 \underset{\mathbf{Q}}{\mathop{\max }}\quad &\frac{1}{S}\sum\limits_{s=1}^{S} {{\log }_{2}}\left| {{\mathbf{I}}_{M_r}}+{\bar P}{{\mathbf{H}}_{eq}}\left( {\mathbf{c}}_{s} \right)\mathbf{QH}_{eq}^{\text{H}}\left( {\mathbf{c}}_{s} \right) \right| \label{eq:calcuRs} \\
 {\rm s.t.}\quad &\frac{1}{K}{{\log }_{2}}\Big| {{\mathbf{I}}_{J}}+{K\bar P\alpha }{{\mathbf{\Psi }}^{\text{H}}}\Big( \mathbf{Q}\otimes {{\left( {{\mathbf{H}}^{\text{H}}}\mathbf{H} \right)}^{\text{T}}} \Big)\mathbf{\Psi } \Big|\ge {{r}_{BD}}, \\
 & {\rm tr}(\mathbf{Q})=1,\mathbf{Q}\succeq0.
\end{align}
(P4) is a convex optimization problem, which can be optimally solved by using software tools like CVX \cite{CVX}.
\subsubsection{Upper Bound Based Solution}
Note that the above sample average based method needs to collect a large number of samples for the random state $\mathbf c(n)$ before solving the stochastic optimization problem. Hence, it requires huge memory to store the samples and the corresponding computational complexity is also high. To address such issues, we substitute \eqref{eq:generalRs} with its upper bound and convert the problem into a more tractable convex problem.
By using Jensen's inequality, an upper bound of $R_s$ in \eqref{eq:generalRs} is given by
\begin{equation}
\begin{split}
  & {{R}_{s}}={{\mathbb{E}}_{\mathbf{c}(n)}}\left[ {{\log }_{2}}\left| {{\mathbf{I}}_{{{M}_{t}}}}+{\bar P}\mathbf{QH}_{eq}^{\text{H}}\left( \mathbf{c}\left( n \right) \right){{\mathbf{H}}_{eq}}\left( \mathbf{c}\left( n \right) \right) \right| \right] \\
 & \le {{R}_{sUB}}\triangleq {{\log }_{2}}\left| {{\mathbf{I}}_{{{M}_{t}}}}+{\bar P}\mathbf{Q}{{\mathbb{E}}_{\mathbf{c}(n)}}\left[ \mathbf{H}_{eq}^{\text{H}}\left( \mathbf{c}\left( n \right) \right){{\mathbf{H}}_{eq}}\left( \mathbf{c}\left( n \right) \right) \right] \right|,\label{eq:Rsupperbound} \\
\end{split}
\end{equation}
where
\begin{small}
\begin{equation}
\begin{split}
  & {{\mathbb{E}}_{\mathbf{c}(n)}}\left[ \mathbf{H}_{eq}^{\text{H}}\left( \mathbf{c}\left( n \right) \right){{\mathbf{H}}_{eq}}\left( \mathbf{c}\left( n \right) \right) \right] \\
 & ={{\mathbb{E}}_{\mathbf{c}(n)}}\Big[ {{\Big( {{\mathbf{H}}_{d}}+\sum\limits_{j=1}^{J}{\sqrt{\alpha }}{{\mathbf{g}}_{j}}\mathbf{h}_{j}^{\text{H}}{{c}_{j}}(n) \Big)}^{\text{H}}}\Big( {{\mathbf{H}}_{d}}+\sum\limits_{j=1}^{J}{\sqrt{\alpha }}{{\mathbf{g}}_{j}}\mathbf{h}_{j}^{\text{H}}{{c}_{j}}(n) \Big) \Big] \\
 & =\mathbf{H}_{d}^{\text{H}}{{\mathbf{H}}_{d}}+\alpha {{\mathbb{E}}_{\mathbf{c}(n)}}\Big[ \sum\limits_{j=1}^{J}{\sum\limits_{i=1}^{J}{c_{j}^{*}(n){{c}_{i}}(n)}}{{\mathbf{h}}_{j}}\mathbf{g}_{j}^{\text{H}}{{\mathbf{g}}_{i}}\mathbf{h}_{i}^{\text{H}} \Big] \\
 & =\mathbf{H}_{d}^{\text{H}}{{\mathbf{H}}_{d}}+\alpha \sum\limits_{j=1}^{J}{{{\left\| {{\mathbf{g}}_{j}} \right\|}^{2}}{{\mathbf{h}}_{j}}}\mathbf{h}_{j}^{\text{H}}.\label{eq:HeqHeq} \\
\end{split}
\end{equation}
\end{small}
The last equality of \eqref{eq:HeqHeq} follows since $c(n)$ are i.i.d. random symbols for different BDs $j$. Therefore, by replacing $R_s$ in \eqref{eq:generalRs} with \eqref{eq:Rsupperbound}, we have
\begin{align}
 \notag(\text{P}5)\quad &\\
 \underset{\mathbf{Q}}{\mathop{\max }}\quad &{{\log }_{2}}\Big| {{\mathbf{I}}_{{{M}_{t}}}}+{\bar P}\mathbf{Q}\Big( \mathbf{H}_{d}^{\text{H}}{{\mathbf{H}}_{d}}+\alpha \sum\limits_{j=1}^{J}{{{\left\| {{\mathbf{g}}_{j}} \right\|}^{2}}{{\mathbf{h}}_{j}}}\mathbf{h}_{j}^{\text{H}} \Big) \Big| \\
 {\rm s.t.}\quad &\frac{1}{K}{{\log }_{2}}\left| {{\mathbf{I}}_{J}}+{K\bar P\alpha }{{\mathbf{\Psi }}^{\text{H}}}\left( \mathbf{Q}\otimes {{\left( {{\mathbf{H}}^{\text{H}}}\mathbf{H} \right)}^{\text{T}}} \right)\mathbf{\Psi } \right|\ge {{r}_{BD}}, \\
 & {\rm tr}(\mathbf{Q})=1,\mathbf{Q}\succeq0.
\end{align}
Apparently, (P5) is a convex optimization problem, which can be optimally solved by using software tools like CVX \cite{CVX}.
\section{Simulation Results}
In this section, simulation results are provided to evaluate the performance of the studied symbiotic radio system. We establish a Cartesian coordinate system, where the PT is located at the origin (0, 0),
%the $J$ BDs are located at the (190m, 10m),
and the $J$ BDs are randomly distributed in a circle with (180m, 20m) as the center and a radius of 5m. Furthermore, the AP is located at (200m, 0). The direct link from the PT to AP is a LoS link, whose channel matrix $\mathbf{H}_{d}$ is generated based on their positions in the Cartesian coordinate system. The links from the PT to BDs and BDs to AP are assumed to be the Rician fading channels with the Rician K-factor $K = 10$ dB. Furthermore, the large-scale channel gains of PT-to-AP and PT-to-BDs links are modeled as ${{\beta }}={{\beta }_{0}}d^{-\gamma }$ , where ${{\beta }_{0}}={{\left( \frac{\lambda }{4\pi } \right)}^{2}}$ is the reference channel gain, with $\lambda$ denoting the wavelength, $d$ represents the corresponding channel link distance of PT-to-AP or PT-to-BDs, and $\gamma$ denotes the path loss exponent. We set the path-loss exponents of the PT-to-AP
link, PT-to-BDs links as $\gamma_{TA}$ = 2 and $\gamma_{TB}$ = 2.7, respectively. Furthermore, the large-scale channel coefficients $\beta_{hg}$ of the cascaded PT-BDs-AP
channels are modeled as $\beta_{hg}=0.01\beta_{h}$, where $\beta_{h}$ represents
the large-scale coefficients of PT-to-BDs channels. The carrier frequency is 3.5GHz, the noise power is $\sigma^2=-110$ dBm, the number of transmit antennas at the PT is $M_t=4$, the number of receive antennas at the AP is $M_r=8$, and the power reflection coefficient is $\alpha =1$. Furthermore, we set the ratio between the symbol duration of the BD symbols and that of the PT symbols as $K=128$. For the sample-average based solution, the number of samples $S$ is set as 1000.
%We assume independent random channels for all communication links, where the small-scale fading components follow i.i.d. CSCG distribution with zero mean and unit-variance, and the large-scale channel gains for the PT-to-AP, PT-to-BD, and BD-to-AP channels are $\beta_{Hd}=-120$ dB, $\beta_h=-110$ dB, and $\beta_g=-20$ dB, respectively, i.e., $\text{vec}\left( {{\mathbf{H}}_{d}} \right)\sim\mathcal{C}\mathcal{N}(\mathbf{0},{{\beta }_{Hd}}{{\mathbf{I}}_{M_rM_t}})$, ${{\mathbf{h}}_{j}}\sim \mathcal{C}\mathcal{N}(0,\beta_h\mathbf I_{M_t})$, and $\mathbf g_j\sim \mathcal{CN}(\mathbf 0, \beta_g \mathbf I_{M_r})$, for $j=1,2,\cdots ,J$. The noise power is $\sigma^2=-110$ dBm, the number of transmit antennas at the PT is $M_t=4$, the number of receive antennas at the AP is $M_r=8$, and the power reflection coefficient is $\alpha =1$. Furthermore, we set the ratio between the symbol duration of the BD symbols and that of the PT symbols as $K=128$, the number of samples $S$ as 1000.

\begin{figure}[!t]
  \centering
  \centerline{\includegraphics[height=3in, width=3.5in]{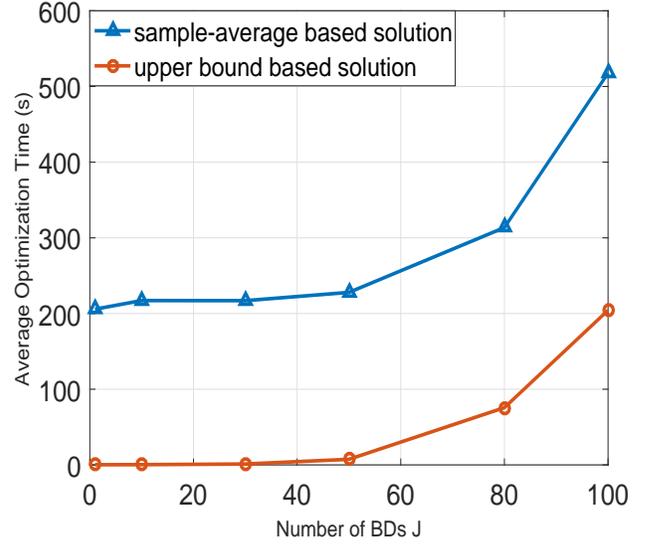}}
  \caption{Average optimization time versus number of BDs $J$.}
  \label{system model}
   \vspace{-0.3cm}
  \end{figure}

\begin{figure}[!t]
  \centering
  \centerline{\includegraphics[height=3in, width=3.5in]{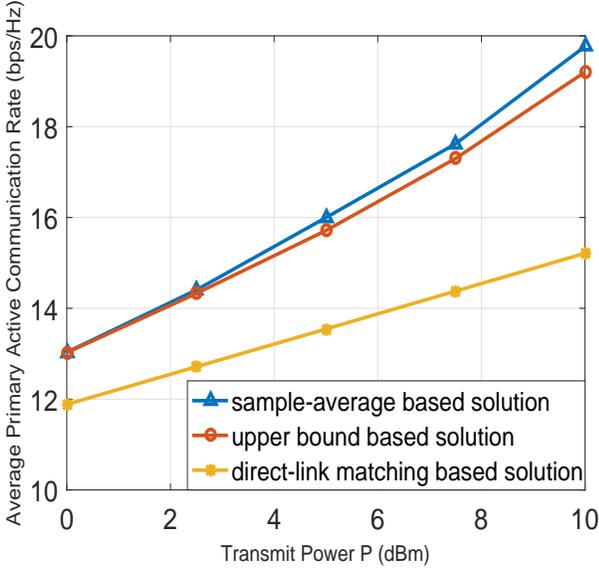}}
  \caption{Average primary active communication rate versus transmit power.}
  \label{system model}
   \vspace{-0.3cm}
  \end{figure}

\begin{figure}[!t]
  \centering
  \centerline{\includegraphics[height=3in, width=3.5in]{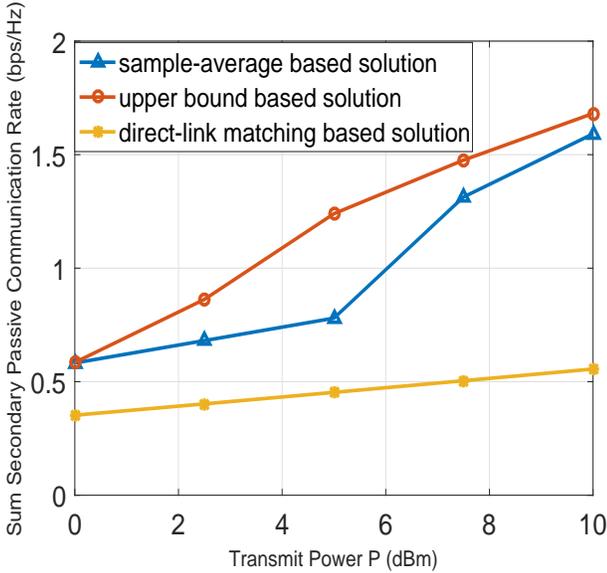}}
  \caption{Secondary passive communication rate versus transmit power.}
  \label{system model}
   \vspace{-0.3cm}
  \end{figure}

Fig. 3 plots the average optimization time versus the number of BDs for the two proposed solutions, where the average is taken over 100 independent channel realizations and $S=1000$ independent samples. The transmit power is $P=0$ dBm. It is observed that the optimization time increases monotonically
with the number of BDs $J$, which is expected since the matrix dimension of the constraint term in \eqref{eq:RBDwithQ} becomes larger as $J$ increases.
It can also be observed that the average optimization time of the upper bound based solution is much lower than that of the sample-average based solution.

Fig. 4 and Fig. 5 compare the performance of the average rate of primary active communication and secondary passive communication with three different precoding schemes, where the average is taken over $S=1000$ independent samples. Besides
our proposed sample average based solution and the upper bound based solution in subsection V B, we also consider the direct-link
matching precoding scheme as a benchmark, where the PT ignores the multipath created by
the BDs and simply sets the precoding matrix to match
the direct link, i.e.,
${\mathbf{F}=\frac{1}{\sqrt{{{M}_{t}}}}{{\mathbf{V}}_{{\mathbf{H}_{d}}}}{{\mathbf{P}}^{\frac{1}{2}}}}$, with
${{\mathbf{V}}_{{\mathbf{H}_{d}}}}$ obtained based on the (reduced) SVD of the channel matrix, ${{\mathbf{{H}}}_{d}}={{\mathbf{U}}_{{{{\mathbf{{H}}}}_{d}}}}{{\mathbf{\Sigma }}_{{{{\mathbf{{H}}}}_{d}}}}\mathbf{V}_{{{{\mathbf{{H}}}}_{d}}}^{\text{H}}$.
%with $\mathbf{\Sigma }_{{{{\mathbf{{H}}}}_{d}}}^{\text{H}}{{\mathbf{\Sigma }}_{{{{\mathbf{{H}}}}_{d}}}}$ a diagonal matrix containing the positive eigenvalues of $\mathbf{{H}}_{d}^{\text{H}}{{\mathbf{{H}}}_{d}}$. ${{\mathbf{U}}_{{{{\mathbf{{H}}}}_{d}}}}\in {{\mathbb{C}}^{{{M}_{r}}\times \text{rank}\left( {{{\mathbf{{H}}}}_{d}} \right)}}$ and ${{\mathbf{V}}_{{{{\mathbf{{H}}}}_{d}}}}\in {{\mathbb{C}}^{{{M}_{t}}\times \text{rank}\left( {{{\mathbf{{H}}}}_{d}} \right)}}$ are both semi-unitary matrixes.
The results in Fig. 4 and Fig. 5 are
obtained for one realization of the channels, and the number
of BDs is $J = 50$. It is observed that for all
the three precoding schemes, the average primary active rate and secondary passive rate increase
monotonically with the transmit power $P$, as expected. Furthermore, both our proposed solutions in subsection V B significantly outperform the benchmarking direct-link matching precoding scheme, thanks to the consideration
of the effective channel constituted by both the direct link and
the backscattered multipaths. Furthermore, it is observed that the sample-average based solution achieves slightly higher primary rate than the upper bound based solution, but at the cost of higher computation complexity, as illustrated in Fig. 3.

\begin{figure}[!t]
  \centering
  \centerline{\includegraphics[height=3in, width=3.5in]{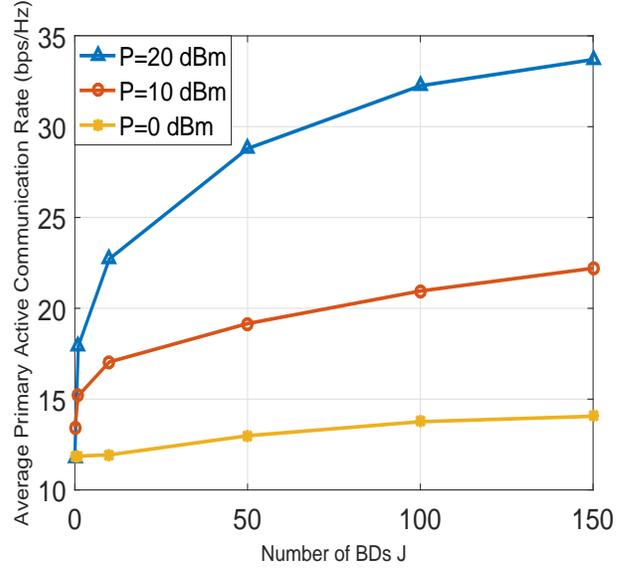}}
  \caption{Average primary communication rate versus number of BDs $J$.}
  \label{system model}
   \vspace{-0.3cm}
  \end{figure}
\begin{figure}[!t]
  \centering
  \centerline{\includegraphics[height=3in, width=3.5in]{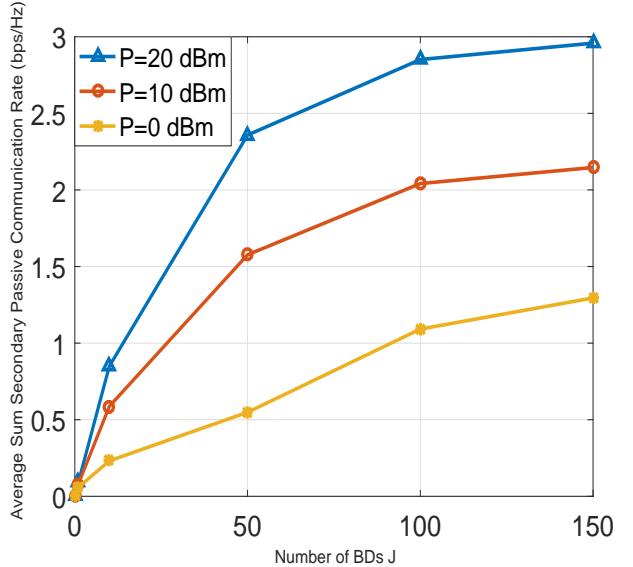}}
  \caption{Average secondary communication rate versus number of BDs $J$.}
  \label{system model}
   \vspace{-0.3cm}
  \end{figure}
Next, we study the impact of the number of BDs $J$ on the average primary and secondary communication rates,
where the average is taken over 100 independent channel realizations and $S$ independent samples.
For each channel realization, the primary communication rate is obtained with the proposed upper bound based solution. Note that though sub-optimal in general, the upper bound based solution has much lower computational complexity than the sample-average based solution.
Fig. 6 and Fig. 7 plot the average primary and secondary communication rates versus the number of BDs $J$, respectively. It is firstly observed that the primary communication rate is in general much higher than the secondary rate. This is expected since the symbol rate of primary signals is $K=128$ times of that of the secondary signals, and that the backscattered link for one single BD is in general much weaker than the primary communication link. Furthermore, it is observed that as $J$ increases, both primary and secondary rates increase, which corroborates our theoretical results in Section III.

\section{Conclusion}
In this paper, a MIMO symbiotic radio system with a massive number of BDs was investigated. The achievable rates of both the primary active communication and secondary passive communication were derived. Furthermore, considering the asymptotic regime as the number
of BDs goes large, closed-form expressions were derived for the general
MIMO symbiotic radio setup and the special SIMO setup, both of which were shown to be increasing functions of the number of BDs.
This thus demonstrated that the mutualism relationship of symbiotic radio can be fully exploited with massive BD access.
In addition, the precoding optimization
problem was studied to maximize the primary communication rate
while guaranteeing that the secondary communication rate was no
smaller than a certain threshold.
Extensive simulation results were provided to verify the effectiveness of our proposed solutions.

\begin{appendices}
\section{Proof of Theorem 1}
Denote the right hand side of \eqref{eq:sumrate2} in Theorem 1 as $C$, which can be expressed as
\begin{equation}
\begin{split}
  C& =\frac{1}{K}{{\log }_{2}}\Big| {{\mathbf{I}}_{M_rM_s}}+\frac{1}{{{\sigma }^{2}}}\sum\limits_{j=1}^{J}{{{\mathbf{x}}_{j}}\mathbf{x}_{j}^{\text{H}}} \Big| \\
 & =\frac{1}{K}{{\log }_{2}}\Big| {{\mathbf{I}}_{M_rM_s}}+\frac{1}{{{\sigma }^{2}}}\sum\limits_{j=2}^{J}{{{\mathbf{x}}_{j}}\mathbf{x}_{j}^{\text{H}}+\frac{1}{{{\sigma }^{2}}}{{\mathbf{x}}_{1}}\mathbf{x}_{1}^{\text{H}}} \Big|. \\
\end{split}
\end{equation}
Let ${{\mathbf{A}}_{i}}={{\mathbf{I}}_{M_rM_s}}+\frac{1}{{{\sigma }^{2}}}\sum\limits_{j=i+1}^{J}{{{\mathbf{x}}_{j}}\mathbf{x}_{j}^{\text{H}}}$, then $C$ can be expressed as
\begin{equation}
\begin{split}
 C& =\frac{1}{K}{{\log }_{2}}\left| {{\mathbf{A}}_{0}} \right|=\frac{1}{K}{{\log }_{2}}\Big| {{\mathbf{A}}_{1}}+\frac{1}{{{\sigma }^{2}}}{{\mathbf{x}}_{1}}\mathbf{x}_{1}^{\text{H}} \Big| \\
 & =\frac{1}{K}{{\log }_{2}}\Big| {{\mathbf{A}}_{1}}\Big( {{\mathbf{I}}_{M_rM_s}}+\frac{1}{{{\sigma }^{2}}}\mathbf{A}_{1}^{-1}{{\mathbf{x}}_{1}}\mathbf{x}_{1}^{\text{H}} \Big) \Big| \\
 & =\frac{1}{K}{{\log }_{2}}\left| {{\mathbf{A}}_{1}} \right|+\frac{1}{K}{{\log }_{2}}\Big| {{\mathbf{I}}_{M_rM_s}}+\frac{1}{{{\sigma }^{2}}}\mathbf{A}_{1}^{-1}{{\mathbf{x}}_{1}}\mathbf{x}_{1}^{\text{H}} \Big| \\
 & =\frac{1}{K}{{\log }_{2}}\left| {{\mathbf{A}}_{1}} \right|+\frac{1}{K}{{\log }_{2}}\Big( 1+\frac{1}{{{\sigma }^{2}}}\mathbf{x}_{1}^{\text{H}}\mathbf{A}_{1}^{-1}{{\mathbf{x}}_{1}} \Big) \\
 & =\frac{1}{K}{{\log }_{2}}\left| {{\mathbf{A}}_{1}} \right|+\frac{1}{K}{{\log }_{2}}\left( 1+{{\gamma }_{{{c}_{1}}}} \right), \\
\end{split}
\end{equation}
where the second last equality follows from the Weinstein-Aronszajn identity $\left| {{\mathbf{I}}_{m}}+\mathbf{AB} \right|=\left| {{\mathbf{I}}_{n}}+\mathbf{BA} \right|$, and ${\gamma }_{{c}_{1}}$ is the maximum SINR of BD 1 given in \eqref{BDjmaximumSINR}. By applying the similar decomposition,
$\frac{1}{K}{{\log }_{2}}\left| {{\mathbf{A}}_{1}} \right|$ can be expressed as
\begin{equation}
{\frac{1}{K}{{\log }_{2}}\left| {{\mathbf{A}}_{1}} \right|=\frac{1}{K}{{\log }_{2}}\left| {{\mathbf{A}}_{2}} \right|+\frac{1}{K}{{\log }_{2}}\left( 1+{{\gamma }_{{{c}_{2}}}} \right)}.
\end{equation}
By applying the above result recursively, $C$ can be expressed as a general form
\begin{equation}
{C=\frac{1}{K}{{\log }_{2}}\left| {{\mathbf{A}}_{j}} \right|+\frac{1}{K}\sum\limits_{i=1}^{j}{{{\log }_{2}}\left( 1+{{\gamma }_{{{c}_{i}}}} \right)}}.\label{generalC}
\end{equation}
By letting $j=J$, \eqref{generalC} can be expressed as
\begin{equation}
\begin{split}
 C& =\frac{1}{K}{{\log }_{2}}\left| {{\mathbf{A}}_{J}} \right|+\frac{1}{K}\sum\limits_{i=1}^{J}{{{\log }_{2}}\left( 1+{{\gamma }_{{{c}_{i}}}} \right)} \\
 & =\frac{1}{K}\sum\limits_{j=1}^{J}{{{\log }_{2}}\left( 1+{{\gamma }_{{{c}_{j}}}} \right)}=R_{BD}. \\
\end{split}
\end{equation}
This thus completes the proof of Theorem 1.
\section{Proof of Lemma 1}
To show Lemma 1, by substituting ${\mathbf{x}}_{j}=\sqrt{KP\alpha}\text{vec}\left( {{\mathbf{g}}_{j}} \mathbf{h}_{j}^{\text{H}}\mathbf{F} \right)$, we have
\begin{equation}
\begin{split}
  {{\mathbf{x}}_{j}}\mathbf{x}_{j}^{\text{H}}& =KP\alpha \text{vec}\left( {{\mathbf{g}}_{j}}\cdot 1\cdot \mathbf{h}_{j}^{\text{H}}\mathbf{F} \right)\text{vec}{{\left( {{\mathbf{g}}_{j}}\cdot 1\cdot \mathbf{h}_{j}^{\text{H}}\mathbf{F} \right)}^{\text{H}}} \\
 & =KP\alpha \left( {{\left( \mathbf{h}_{j}^{\text{H}}\mathbf{F} \right)}^{\text{T}}}\otimes {{\mathbf{g}}_{j}} \right){{\left( {{\left( \mathbf{h}_{j}^{\text{H}}\mathbf{F} \right)}^{\text{T}}}\otimes {{\mathbf{g}}_{j}} \right)}^{\text{H}}} \\
 & =KP\alpha \left( {{\mathbf{F}}^{\text{T}}}\mathbf{h}_{j}^{*}\mathbf{h}_{j}^{\text{T}}{{\mathbf{F}}^{*}} \right)\otimes \left( {{\mathbf{g}}_{j}}\mathbf{g}_{j}^{\text{H}} \right),
\end{split}
\end{equation}
where the second equality follows from the identity $\text{vec}({{\mathbf{A}}_{1}}{{\mathbf{A}}_{2}}{{\mathbf{A}}_{3}})=\left( \mathbf{A}_{3}^{\text{T}}\otimes {{\mathbf{A}}_{1}} \right)\text{vec}({{\mathbf{A}}_{2}})$, and the last equality follows from ${{\left( \mathbf{A}\otimes \mathbf{B} \right)}^{\text{H}}}={{\mathbf{A}}^{\text{H}}}\otimes {{\mathbf{B}}^{\text{H}}}$ and $\left( \mathbf{A}\otimes \mathbf{B} \right)\left( \mathbf{C}\otimes \mathbf{D} \right)=\mathbf{AC}\otimes \mathbf{BD}$.
Thus, ${{R}_{BD}}$ in \eqref{eq:sumrate2} can be expressed as
\begin{small}
\begin{equation}
\begin{split}
  {{R}_{BD}}& =\frac{1}{K}{{\log }_{2}}\Big| {{\mathbf{I}}_{M_rM_s}}+\frac{1}{{{\sigma }^{2}}}\sum\limits_{j=1}^{J}{{{\mathbf{x}}_{j}}\mathbf{x}_{j}^{\text{H}}} \Big| \\
 & =\frac{1}{K}{{\log }_{2}}\Big| {{\mathbf{I}}_{M_rM_s}}+{K\bar P\alpha }\sum\limits_{j=1}^{J}{\left( \left( {{\mathbf{F}}^{\text{T}}}\mathbf{h}_{j}^{*}\mathbf{h}_{j}^{\text{T}}{{\mathbf{F}}^{*}} \right)\otimes {{\mathbf{g}}_{j}}\mathbf{g}_{j}^{\text{H}} \right)} \Big| \\
 & =\frac{1}{K}{{\log }_{2}}\Big| {{\Big( {{\mathbf{I}}_{M_rM_s}}+{K\bar P\alpha }\sum\limits_{j=1}^{J}{\left( \left( {{\mathbf{F}}^{\text{T}}}\mathbf{h}_{j}^{*}\mathbf{h}_{j}^{\text{T}}{{\mathbf{F}}^{*}} \right)\otimes {{\mathbf{g}}_{j}}\mathbf{g}_{j}^{\text{H}} \right)} \Big)}^{\text{T}}} \Big| \\
 & =\frac{1}{K}{{\log }_{2}}\Big| {{\mathbf{I}}_{M_rM_s}}+{K\bar P\alpha }\sum\limits_{j=1}^{J}{\left( \left( {{\mathbf{F}}^{\text{H}}}{{\mathbf{h}}_{j}}\mathbf{h}_{j}^{\text{H}}\mathbf{F} \right)\otimes {{\left( {{\mathbf{g}}_{j}}\mathbf{g}_{j}^{\text{H}} \right)}^{\text{T}}} \right)} \Big|.
\end{split}
\end{equation}
\end{small}
This thus completes the proof of Lemma 1.
\section{Proof of Lemma 2}
By using the law of large numbers, for $J\gg 1$, we have:
\begin{equation}
\begin{split}
  & \sum\limits_{j=1}^{J}{\left( \left( {{\mathbf{F}}^{\text{H}}}{{\mathbf{h}}_{j}}\mathbf{h}_{j}^{\text{H}}\mathbf{F} \right)\otimes {{\left( {{\mathbf{g}}_{j}}\mathbf{g}_{j}^{\text{H}} \right)}^{\text{T}}} \right)} \\
 & \to J\mathbb{E}\left[ \left( {{\mathbf{F}}^{\text{H}}}{{\mathbf{h}}_{j}}\mathbf{h}_{j}^{\text{H}}\mathbf{F} \right)\otimes {{\left( {{\mathbf{g}}_{j}}\mathbf{g}_{j}^{\text{H}} \right)}^{\text{T}}} \right] \\
 & =J\mathbb{E}\left[ \left( {{\mathbf{F}}^{\text{H}}}{{\mathbf{h}}_{j}}\mathbf{h}_{j}^{\text{H}}\mathbf{F} \right) \right]\otimes \mathbb{E}\left[ {{\left( {{\mathbf{g}}_{j}}\mathbf{g}_{j}^{\text{H}} \right)}^{\text{T}}} \right] \\
 & =J{{\beta }_{h}}{{\beta }_{g}}\left( {{\mathbf{F}}^{\text{H}}}\mathbf{F} \right)\otimes {{\mathbf{I}}_{M_r}}.
\end{split}
\end{equation}
It then follows from \eqref{eq:RBDtrans} that
\begin{equation}
\begin{split}
 {{R}_{BD}}& \to \frac{1}{K}{{\log }_{2}}\left| {{\mathbf{I}}_{M_rM_s}}+{JK\bar P\alpha {{\beta }_{h}}{{\beta }_{g}}}\left( {{\mathbf{F}}^{\text{H}}}\mathbf{F}\otimes {{\mathbf{I}}_{M_r}} \right) \right| \\
 & =\frac{1}{K}{{\log }_{2}}\left| \left( {{\mathbf{I}}_{M_s}}+{JK\bar P\alpha {{\beta }_{h}}{{\beta }_{g}}}{{\mathbf{F}}^{\text{H}}}\mathbf{F} \right)\otimes {{\mathbf{I}}_{M_r}} \right| \\
 & =\frac{1}{K}{{\log }_{2}}\left( {{\left| {{\mathbf{I}}_{M_s}}+{JK\bar P\alpha {{\beta }_{h}}{{\beta }_{g}}}{{\mathbf{F}}^{\text{H}}}\mathbf{F} \right|}^{M_r}}{{\left| {{\mathbf{I}}_{M_r}} \right|}^{M_s}} \right) \\
 & =\frac{M_r}{K}{{\log }_{2}}\left| {{\mathbf{I}}_{M_s}}+{JK\bar P\alpha {{\beta }_{h}}{{\beta }_{g}}}{{\mathbf{F}}^{\text{H}}}\mathbf{F} \right|\\
 & =\frac{M_r}{K}{{\log }_{2}}\left| {{\mathbf{I}}_{M_t}}+{JK\bar P\alpha {{\beta }_{h}}{{\beta }_{g}}}\mathbf{F}{{\mathbf{F}}^{\text{H}}} \right|,
\end{split}\label{eq:RBDJ>>1L>1}
\end{equation}
where the third last equality follows from the identity $\left| \mathbf{A}\otimes \mathbf{B} \right|={{\left| \mathbf{A} \right|}^{\text{rank}(\mathbf{B})}}{{\left| \mathbf{B} \right|}^{\text{rank}(\mathbf{A})}}$.

This completes the proof of Lemma 2.
\section{Proof of Lemma 3}
According to \eqref{eq:primaryrate}, we have
\begin{equation}
\begin{split}
  {{R}_{s}}& ={{\mathbb{E}}_{\mathbf{c}(n)}}\left[ {{\log }_{2}}\left| {{\mathbf{I}}_{{{M}_{r}}}}+{\bar P}{{\mathbf{H}}_{eq}}\left( \mathbf{c}\left( n \right) \right)\mathbf{F}{{\mathbf{F}}^{\text{H}}}\mathbf{H}_{eq}^{\text{H}}\left( \mathbf{c}\left( n \right) \right) \right| \right] \\
 & ={{\mathbb{E}}_{\mathbf{c}(n)}}\left[ {{\log }_{2}}\left| {{\mathbf{I}}_{{{M}_{t}}}}+{\bar P}\mathbf{F}{{\mathbf{F}}^{\text{H}}}\mathbf{H}_{eq}^{\text{H}}\left( \mathbf{c}\left( n \right) \right){{\mathbf{H}}_{eq}}\left( \mathbf{c}\left( n \right) \right) \right| \right].\label{eq:primaryratetransform} \\
\end{split}
\end{equation}
\begin{small}
Furthermore, we have
\begin{equation}
\begin{split}
  & \mathbf{H}_{eq}^{\text{H}}(\mathbf{c}(n)){{\mathbf{H}}_{eq}}(\mathbf{c}(n)) \\
 & ={{\Big( {{\mathbf{H}}_{d}}+\sum\limits_{j=1}^{J}{\sqrt{\alpha }}{{\mathbf{g}}_{j}}\mathbf{h}_{j}^{\text{H}}{{c}_{j}}(n) \Big)}^{\text{H}}}\Big( {{\mathbf{H}}_{d}}+\sum\limits_{j=1}^{J}{\sqrt{\alpha }}{{\mathbf{g}}_{j}}\mathbf{h}_{j}^{\text{H}}{{c}_{j}}(n) \Big) \\
 & =\mathbf{H}_{d}^{\text{H}}{{\mathbf{H}}_{d}}+\sqrt{\alpha }\sum\limits_{j=1}^{J}{c_{j}^{*}(n)}{{\mathbf{h}}_{j}}\mathbf{g}_{j}^{\text{H}}{{\mathbf{H}}_{d}}+\sqrt{\alpha }\sum\limits_{j=1}^{J}{{{c}_{j}}(n)\mathbf{H}_{d}^{\text{H}}}{{\mathbf{g}}_{j}}\mathbf{h}_{j}^{\text{H}} \\
 & +\alpha \sum\limits_{j=1}^{J}{\sum\limits_{i=1}^{J}{c_{j}^{*}(n){{c}_{i}}(n)}}{{\mathbf{h}}_{j}}\mathbf{g}_{j}^{\text{H}}{{\mathbf{g}}_{i}}\mathbf{h}_{i}^{\text{H}} \\
 & =\mathbf{H}_{d}^{\text{H}}{{\mathbf{H}}_{d}}+\sqrt{\alpha }\sum\limits_{j=1}^{J}{c_{j}^{*}(n)}{{\mathbf{h}}_{j}}\mathbf{g}_{j}^{\text{H}}{{\mathbf{H}}_{d}}+\sqrt{\alpha }\sum\limits_{j=1}^{J}{{{c}_{j}}(n)\mathbf{H}_{d}^{\text{H}}}{{\mathbf{g}}_{j}}\mathbf{h}_{j}^{\text{H}} \\
 & +\alpha \sum\limits_{j=1}^{J}{{{\left| {{c}_{j}}(n) \right|}^{2}}}{{\left\| {{\mathbf{g}}_{j}} \right\|}^{2}}{{\mathbf{h}}_{j}}\mathbf{h}_{j}^{\text{H}}+\alpha \sum\limits_{j=1,j\ne i}^{J}{\sum\limits_{i=1}^{J}{c_{j}^{*}(n){{c}_{i}}(n)}}{{\mathbf{h}}_{j}}\mathbf{g}_{j}^{\text{H}}{{\mathbf{g}}_{i}}\mathbf{h}_{i}^{\text{H}}. \\
\end{split}
\end{equation}
\end{small}
Due to the law of large numbers, for $J\gg 1$, we have:
$\alpha \sum\limits_{j=1,j\ne i}^{J}{\sum\limits_{i=1}^{J}{c_{j}^{*}(n){{c}_{i}}(n)}}
{{\mathbf{h}}_{j}}\mathbf{g}_{j}^{\text{H}}{{\mathbf{g}}_{i}}\mathbf{h}_{i}^{\text{H}}\to 0$. Therefore,
\begin{equation}
\begin{split}
  & \mathbf{H}_{eq}^{\text{H}}(\mathbf{c}(n)){{\mathbf{H}}_{eq}}(\mathbf{c}(n)) \\
 & \to \mathbf{H}_{d}^{\text{H}}{{\mathbf{H}}_{d}}+J\sqrt{\alpha }\mathbb{E}\left[ c_{j}^{*}(n){{\mathbf{h}}_{j}}\mathbf{g}_{j}^{\text{H}}{{\mathbf{H}}_{d}} \right]+J\sqrt{\alpha }\mathbb{E}\left[ {{c}_{j}}(n)\mathbf{H}_{d}^{\text{H}}{{\mathbf{g}}_{j}}\mathbf{h}_{j}^{\text{H}} \right] \\
 & +J\alpha \mathbb{E}\left[ {{\left| {{c}_{j}}(n) \right|}^{2}}{{\left\| {{\mathbf{g}}_{j}} \right\|}^{2}}{{\mathbf{h}}_{j}}\mathbf{h}_{j}^{\text{H}} \right] \\
 & =\mathbf{H}_{d}^{\text{H}}{{\mathbf{H}}_{d}}+J\alpha {{M}_{r}}{{\beta }_{g}}{{\beta }_{h}}{{\mathbf{I}}_{{{M}_{t}}}}. \label{HeqHeq}\\
\end{split}
\end{equation}
By substituting \eqref{HeqHeq} into \eqref{eq:primaryratetransform}, the proof of Lemma 3 thus completes.

\end{appendices}


%\end{IEEEbiography}

%\newpage



%\vfill


\end{document}